\documentclass{an}

\usepackage{graphicx}
\usepackage{times}
\usepackage{natbib}

\usepackage{graphicx}
\usepackage{txfonts}
\usepackage{epsfig}
\usepackage{times}
\usepackage{rotating}
\usepackage{setspace}

\def\gtsim{\mathrel{\spose{\lower.5ex \hbox{$\mathchar"218$}}
     \raise.4ex\hbox{$\mathchar"13E$}}}
\def\ltsim{\mathrel{\spose{\lower.5ex\hbox{$\mathchar"218$}}
     \raise.4ex\hbox{$\mathchar"13C$}}}
\def\aFe{[$\alpha/{\rm Fe}$]~}

\def\Hb{${\rm H}_{\beta}$}

\def\Mgb{{\rm Mg}\,$_b$}
\def\Fe{$\langle {\rm Fe}\rangle$}

\def\ZH{[$Z/{\rm H}$]~}
\def\MgFe{[${\rm MgFe}$]$'$}

\def\Mgd{{\rm Mg}\,$_2$}

\def\Rbd{$r_{\rm{ bd}}$}

\def\kms{$\rm km\;s^{-1}$}

\def\spose#1{\hbox to 0pt{#1\hss}}

\def\aj{AJ}                   
\def\araa{ARA\&A}             
\def\apj{ApJ}                 
\def\apjs{ApJS}               
\def\aap{A\&A}                
\def\mnras{MNRAS}             

\def\kms{$\rm km\;s^{-1}$}

\def\Hb{${\rm H}{\small{\beta}}$}

\begin{document}

\Pagespan{0}{}
\Yearpublication{}%
\Yearsubmission{2005}%
\Month{11}%
\Volume{999}%
\Issue{88}%

\title{Stellar populations of the bulges of four spiral galaxies,\thanks{Based on observations made with ESO Telescopes at the La Silla-Paranal Observatory under programmes 075.B-0794 and 077.B-0767.},\thanks{Table~2 is available in electronic form.}}

\author{L. Morelli\inst{1,2}\fnmsep\thanks{Corresponding author:
  \email{lorenzo.morelli@unipd.it}\newline}
\and   A. Pizzella\inst{1,2}
\and   E. M. Corsini\inst{1,2}
\and   E. Dalla Bont\`a\inst{1,2}
\and   L. Coccato\inst{3}
\and   J.~M\'endez-Abreu\inst{4}
\and   M. Parmiggiani\inst{2}
}
\titlerunning{Stellar populations of the bulges}
\authorrunning{Morelli L. et al.}
\institute{
Dipartimento di Fisica e Astronomia ``G. Galilei", Universit\`a di Padova,
vicolo dell'Osservatorio 3, I-35122 Padova, Italy
\and 
INAF-Osservatorio Astronomico di Padova, vicolo dell'Osservatorio~2, I-35122 Padova, Italy
\and 
European Southern Observatory, Karl-Schwarzschild-Stra$\beta$e 2, D-85748 Garching bei M\"unchen, Germany. 
\and 
School of Physics and Astronomy, University of St. Andrews, North Haugh, 
St. Andrews KY16 9SS, UK}

\received{}
\accepted{}
\publonline{later}


\abstract{%
Key information to understand the formation and evolution of disk
galaxies are imprinted in the stellar populations of their bulges. This
paper has the purpose to make available new measurements of the
stellar population properties of the bulges of four spiral
galaxies. Both the central values and radial profiles of the line
strength of some of the most common Lick indices are measured along
the major- and minor- axis of the bulge-dominated region of the sample
galaxies. The corresponding age, metallicity, and $\alpha/$Fe ratio
are derived by using the simple stellar population synthesis model
predictions.  The central values and the gradients of the stellar
population properties of ESO-LV~1890070, ESO-LV~4460170, and
ESO-LV~5140100 are consistent with previous findings for bulges of
spiral galaxies. On the contrary, the bulge of ESO-LV 4500200 shows
peculiar chemical properties possibly due to the presence of a central
kinematically-decoupled component. The negative metallicity gradient
found in our bulges sample indicates a relevant role for the
dissipative collapse in bulge formation. However, the shallow
gradients found for the age and $\alpha/$Fe ratio suggests that
merging can not be completely ruled out for the sample bulges. This is
confirmed by the properties of ESO-LV~4500200 which can hardly be
explained without invoking the capture of external material.
}

\maketitle

\section{Introduction}
\label{sec:introduction}

Due to their privileged position at the bottom of the galactic
potential well, bulges are a key player in the process of the assembly
of disk galaxies.
In the current picture, the mechanisms of bulge formation include
dissipative collapse \citep{gilwys98} merging and acquisition events
\citep{coletal00}, and secular evolution \citep{korken04}.
Crucial information to understand the processes of formation and
evolution of galaxies is imprinted in their stellar populations and
even more in their radial gradients, since different formation
scenarios predict different radial trends of age, metallicity, and
$\alpha/$Fe ratio.

In the last years the stellar populations of classical bulges have
been spectroscopically studied in detail and compared  among
different morphological types and different environments in both
lenticular \citep{rampetal05, sancetal06p, colletal06, annietal07,
  rawletal10, kuntetal10, spoletal10} and spiral galaxies
\citep{mooretal06, jabletal07, moreetal08, macaetal09, morelli2012,
  moreetal13}.
The central values and radial gradients of age, metallicity, and
$\alpha/$Fe ratio have been derived for a large number of
galaxies. The variety of the results testifies the complexity of the
topic. The bulk of the stellar population can have a range of
  ages among different bulges and this is generally related with the
morphological type \citep{gandetal07}. The values of the $\alpha/$Fe
 ratio, ranging from solar to super-solar, give a time-scale for
star formation from 4-5 Gyr to less than 1 Gyr \citep{thda06}. The
only common feature for all the bulges (independently from their
morphological type and environment) is the negative radial gradient of
metallicity. All these properties support formation scenarios related
to an early formation through violent relaxation or dissipative
collapse are favoured over those of secular evolution.

To further investigate this topic, \citet{morelli2012} analyzed a
sample of bulges embedded in low surface-brightness disks. Low
surface-brightness galaxies are believed to not have experienced major
merging events during their lifetime. The radial profiles of their
age, metallicity, and $\alpha/$Fe ratio confirm also for these bulges
the violent relaxation as possible formation mechanism. Furthermore,
the comparison between ordinary high and low surface-brightness
galaxies shows that their bulges share many structural and chemical
properties. Such similarity suggests that they possibly had common
formation scenarios and evolution histories and indicates that there
is not a relevant interplay between the bulge and disk components. All
these results downsize the role of the secular evolution in the
formation scenarios of the classical bulges, independently of the high
or low surface-brightness of the host disk. However, secular evolution
could still be the main mechanism responsible for the formation of
pseudobulges \citep{kormetal09}.

In this paper, we present new measurements of the central
values and radial profiles of the line strength of some of the most
common Lick indices for the bulges hosted by four spiral
galaxies. These data are a valuable supplementary resource for the
astronomical community. Indeed, the derived values of age,
metallicity, and $\alpha/$Fe ratio can be used for further comparison
between the stellar populations of galactic components.

\begin{table*}
\caption{Properties of the sample galaxies.}
\begin{center}
\begin{small}
\begin{tabular}{lllr cr ccc}
\hline
\noalign{\smallskip}
\multicolumn{1}{c}{Name} &
\multicolumn{1}{c}{Alt. Name} &
\multicolumn{1}{c}{Type} &
\multicolumn{1}{c}{T} &
\multicolumn{1}{c}{$D_{25}\times d_{25}$} &
\multicolumn{1}{c}{$B_T$} &
\multicolumn{1}{c}{$V_{\rm CMB}$} &
\multicolumn{1}{c}{$D$} &
\multicolumn{1}{c}{$M_{B_T}$} \\
\noalign{\smallskip}
\multicolumn{1}{c}{} &
\multicolumn{1}{c}{} &
\multicolumn{1}{c}{} &
\multicolumn{1}{c}{} &
\multicolumn{1}{c}{[arcmin]} &
\multicolumn{1}{c}{[mag]} &
\multicolumn{1}{c}{[\kms]} &
\multicolumn{1}{c}{[Mpc]} &
\multicolumn{1}{c}{[mag]} \\
\noalign{\smallskip}
\multicolumn{1}{c}{(1)} &
\multicolumn{1}{c}{(2)} &
\multicolumn{1}{c}{(3)} &
\multicolumn{1}{c}{(4)} &
\multicolumn{1}{c}{(5)} &
\multicolumn{1}{c}{(6)} &
\multicolumn{1}{c}{(7)} &
\multicolumn{1}{c}{(8)} &
\multicolumn{1}{c}{(9)} \\
\noalign{\smallskip}
\hline
\noalign{\smallskip}  
ESO-LV~1890070 &  NGC~7140 & SABb & $3.8$ & $3.0\times2.0$ &12.31 & 2981 & 37.5 & $-20.56$ \\
ESO-LV~4460170 &  ...      & SBb  & $3.3$ & $2.2\times1.5$ &13.55 & 4172 & 58.9 & $-20.30$ \\
ESO-LV~4500200 &  NGC~6000 & SBbc & $4.1$ & $1.9\times1.6$ &13.03 & 2118 & 31.6 & $-19.47$ \\
ESO-LV~5140100 &  IC~4538  & SABc & $5.1$ & $2.5\times1.9$ &12.88 & 2888 & 40.4 & $-20.13$ \\
\noalign{\smallskip}
\hline
\noalign{\medskip}
\end{tabular}
\end{small}
\label{tab:sample}
\end{center}
\begin{minipage}{17.1cm}
\begin{small}
NOTES: Col.(3): morphological classification from Lyon Extragalactic
Database (LEDA). Col.(4): morphological type code from LEDA.  Col.(5):
apparent isophotal diameters measured at a surface-brightness level of
$\mu_B = 25$ mag arcsec$^{-2}$ from LEDA. Col.(6): total observed blue
magnitude from LEDA. Col.(7): radial velocity with respect to the CMB
radiation from LEDA. Col.(8): distance from \citet{pizzetal08}
adopting $H_0 = 75$ km s$^{-1}$ Mpc$^{-1}$. Col.(9): absolute total
blue magnitude from $B_T$ corrected for extinction as in LEDA and
adopting $D$.
\end{small}
\end{minipage}
\end{table*}

\section{Galaxy sample}

The galaxy sample comprises four spiral galaxies whose basic
properties are listed in Table~\ref{tab:sample}.  They belong to the
sample of nearby galaxies studied by \citet{pizzetal08} who analyzed
their stellar and ionized-gas kinematics and surface
photometry. \citet{pizzetal08} were interested in spiral galaxies with
a low surface-brightness disk and we refer to their paper for the
details about the selection criteria of the galaxy sample.
ESO-LV~1890070, ESO-LV~4460170, ESO-LV~4500200, and ESO-LV~5140100
turned out to host a high surface-brightness disk and therefore
were not considered in the subsequent analysis
\citep{morelli2012}. Since the long-slit spectra of the four neglected
galaxies were available to us, we decided to derive the stellar
population properties of their bulges.

\section{Photometric decomposition}
\label{sec:decomposition}

In their photometric analysis \citet{pizzetal08} assumed the
surface-brightness distribution of the sample galaxies to be the sum
of the contributions of a bulge and a disk component only. We improved
their photometric decomposition by including a bar component to
precisely identify the bulge-dominated region of each galaxy. The
structural parameters of the bulge, disk, and bar components were
derived by applying the Galaxy Surface Photometry Two-Dimensional
Decomposition (GASP2D) algorithm \citep{mendetal08, mendetal14} to the
images obtained by \citet{pizzetal08}. Other components (e.g., lenses,
ovals, or spiral arms) were not considered.

The surface brightness of the bulge was modelled using a S\'ersic
function \citep{sersic68}
\begin{equation}
I_{\rm bulge}(r) = I_{\rm e} 10^{-b_n \left[(r/r_{\rm e})^{1/n}-1\right]},
\end{equation}
where $r_{\rm e}$ is the effective radius, $I_{\rm e}$ is the surface
brightness at $r_{\rm e}$, $n$ is a shape parameter that describes the
curvature of the radial profile, and $b_n = 0.868\,n-0.142$
\citep{caonetal93}. The bulge model was assumed to have elliptical
isophotes centered on $(x_0,y_0)$ with constant position angle 
PA$_{\rm bulge}$ and constant axial ratio $q_{\rm bulge}$.

The surface brightness of the disk was modelled using an exponential
function \citep{freeman70}
\begin{equation}
I_{\rm disk}(r) = I_0 e^{-r/h},
\end{equation}
where $I_0$ is the central surface brightness and $h$ is the scale
length. The disk model was assumed to have elliptical isophotes
centered on $(x_0,y_0)$ with constant position angle PA$_{\rm disk}$ and
constant axial ratio $q_{\rm disk}$.

The surface brightness of the bar was modelled with a Ferrers function
(\citealt{ferrers77}, but see \citealt{aguetal09} for the choice of
the shape parameter)
\begin{equation}
I_{\rm bar}(r) = I_{\rm 0,bar}
\left[1-\left( \frac{r_{\rm bar}}{a_{\rm bar}} \right)^2 \right]^{2.5}
\qquad r_{\rm bar} \leq a_{\rm bar},
\end{equation}
where $I_{0,bar}$ is the central surface brightness and $a_{\rm bar}$
is the bar length. The bar model was assumed to have isophotes
described by generalized ellipses \citep{athetal90} centered on
$(x_0,y_0)$ with constant position angle PA$_{\rm bar}$ and constant
axial ratio $q_{\rm bar}$.

The GASP2D software yields the structural parameters for the bulge
($I_{\rm e}$, $r_{\rm e}$, $n$, PA$_{\rm bulge}$ and $q_{\rm bulge}$),
disk ($I_0$, $h$, PA$_{\rm disk}$ and $q_{\rm disk}$), and bar
($I_{\rm 0, bar}$, $a_{\rm bar}$, PA$_{\rm bar}$ and $q_{\rm bar}$)
and the position of the galaxy center $(x_0, y_0)$ by iteratively
fitting a model of the surface-brightness distribution to the pixels
of the galaxy image. It was used a non-linear least-squares
minimisation based on a robust Levenberg-Marquardt method
\citep{moretal80}. The actual computation has been done using the
MPFIT algorithm \citep{markwardt09} under the IDL\footnote{Interactive
  Data Language is distributed by ITT Visual Information Solutions. It
  is available from http://www.ittvis.com} environment. Each image
pixel has been weighted according to the variance of its total
observed photon counts due to the contribution of both the galaxy and
sky, and determined assuming photon noise limitation and taking into
account the detector readout noise. The seeing effects were taken into
account by convolving the model image with a circular Moffat point
spread function with the shape parameters measured directly from stars
in the galaxy image.

The best-fitting solution was found by building
surface-brightness models with and without the bar component. A bar
was detected and modelled in ESO-LV 4460170 and ESO-LV 4500200,
whereas no bar was needed to improve the surface-brightness models of
ESO-LV 1890070 and ESO-LV 5140100. The bump in the surface-brightness
profile associated to both a significant peak in the ellipticity and a
roughly constant position angle is the photometric signature of the
bar in ESO-LV 4460170 and ESO-LV 4500200 (see \citet{aguetal09} for
a detailed discussion). The flat surface-brightness profile of ESO-LV
1890070 and the increase of ellipticity measured between 30 and 60
arcsec are due to the two prominent and symmetrical arms characterizing
the galaxy spiral pattern. The multi-armed structure of ESO-LV 5140100
results in the increase and variation of the ellipticity between 20
and 40 arcsec.

Figure~\ref{fig:gasp2d} shows the GASP2D fits for the
sample galaxies. The best-fitting parameters derived for their
structural components are collected in Table~\ref{tab:parameters}
together with the radius of the bulge-dominated region where half of
the total surface brightness is due to the bulge only. 

The errors on the best-fitting parameters of the barred galaxies were
obtained through a series of Monte Carlo simulations. A set of 500
images of galaxies with a S\'ersic bulge, an exponential disk, and a
Ferrers bar was generated. The structural parameters of the artificial
galaxies were randomly chosen among the ranges obtained for our
galaxies. The adopted pixel scale, CCD gain, and read-out-noise were
chosen to mimic the instrumental setup of the photometric observations
by \citet{pizzetal08}. A background level and photon noise were added
to the artificial images to yield a signal-to-noise ratio ($S/N$)
similar to that of the observed ones. Finally, the images of
artificial galaxies were analyzed with GASP2D as if they were real and
the errors on the fitted parameters were estimated by comparing the
input and measured values assuming they were normally distributed. The
mean and standard deviation of the relative errors of the artificial
galaxies were adopted as the systematic and typical errors for the
observed galaxies.

\begin{table*}
\caption{Structural parameters of the sample galaxies.}   
\begin{center}
\begin{small}
\begin{tabular}{lcccc}      
\hline
\noalign{\smallskip}
\multicolumn{1}{c}{Parameter} &
\multicolumn{1}{c}{ESO-LV~1890070} &
\multicolumn{1}{c}{ESO-LV~4460170} &
\multicolumn{1}{c}{ESO-LV~4500200} &
\multicolumn{1}{c}{ESO-LV~5140100} \\
\noalign{\smallskip}
\multicolumn{1}{c}{(1)} &
\multicolumn{1}{c}{(2)} &
\multicolumn{1}{c}{(3)} &
\multicolumn{1}{c}{(4)} &
\multicolumn{1}{c}{(5)} \\
\noalign{\smallskip}
\hline
\noalign{\smallskip}  
$\mu_{\rm e}$ [mag arcsec$^{-2}$]       & $18.31\pm 0.13$ & $18.03\pm0.13$ & $16.31\pm0.13$ & $19.2\pm0.07$ \\
$r_{\rm e}$ [arcsec]                    & $3.1\pm0.3$     & $2.6\pm0.3$    & $1.7\pm0.2$  & $1.9\pm0.1$\\
$n$                                     & $1.14\pm0.17$   & $1.28\pm0.19$  & $0.98\pm0.15$  & $1.00\pm0.10$ \\
$q_{\rm bulge}$                         & $0.69\pm0.06 $  & $0.88\pm0.08$  & $0.72\pm0.06$  & $0.89\pm0.04$ \\
PA$_{\rm bulge}$ [$^{\circ}$]           & $16.4\pm1.6$         & $148.7\pm14.9$       & $134.9\pm13.5$       & $43.4\pm4.3$ \\ 
$\mu_0$ [mag arcsec$^{-2}$]             & $19.54\pm0.11$  & $19.70\pm0.11$ & $18.46\pm0.12$    & $19.64\pm0.05$ \\
$h$ [arcsec]                            & $31.2\pm3.1$   & $20.1\pm2.0$   & $18.1\pm1.8$   & $25.2\pm1.5 $ \\
$q_{\rm disk}$                          & $0.45\pm 0.04$  & $0.58\pm0.06$  & $0.71\pm0.07$   & $0.81\pm 0.03$ \\
PA$_{\rm disk}$ [$^{\circ}$]            & $21.2\pm2.1$       & $153.6\pm15.4$       & $155.4\pm15.5$       & $45.2\pm4.5$ \\ 
$\mu_{\rm 0,bar}$ [mag arcsec$^{-2}$]   & $\cdots$        & $20.68\pm0.15$ & $17.97\pm0.15$    & $\cdots$ \\ 
$a_{\rm bar}$ [arcsec]                  & $\cdots$        & $49.1\pm7.4$   & $25.4\pm3.8$       & $\cdots$ \\
$q_{\rm bar}$                           & $\cdots$        & $0.26\pm0.03$  & $0.32\pm0.03$  & $\cdots$ \\
PA$_{\rm bar}$ [$^{\circ}$]             & $\cdots$        & $160.6\pm16.1$       & $172.7\pm17.3$       & $\cdots$ \\
$L_{\rm bulge}/L_{\rm T}$               & $0.01$          & $0.18$         & $0.09$         & $0.02$   \\
$L_{\rm bar}/L_{\rm T}$                 & $\cdots$        & $0.11$         & $0.15$         & $\cdots$ \\
$r_{\rm bd}$ [arcsec]                   & $5.60$          & $4.65$         & $2.60$         & $2.50$   \\   
\noalign{\smallskip}
\hline
\noalign{\smallskip}
\end{tabular}
\end{small}
\label{tab:parameters}
\end{center}
\begin{minipage}{17.1cm}
\begin{small}
NOTES. Surface-brightness values are given in $R$ band according to
the flux calibration by \citet{pizzetal08}.  $L_{\rm bulge}/L_{\rm
  T}$ is the bulge-to-total luminosity ratio.  $L_{\rm bar}/L_{\rm T}$
is the bar-to-total luminosity ratio. $r_{\rm bd}$ is radius of the
bulge-dominated region where the bulge contributes half of the total
surface brightness. 
\end{small}
\end{minipage}
\end{table*}

\begin{figure*}
\centering
   \includegraphics[angle=0,width=0.9\textwidth]{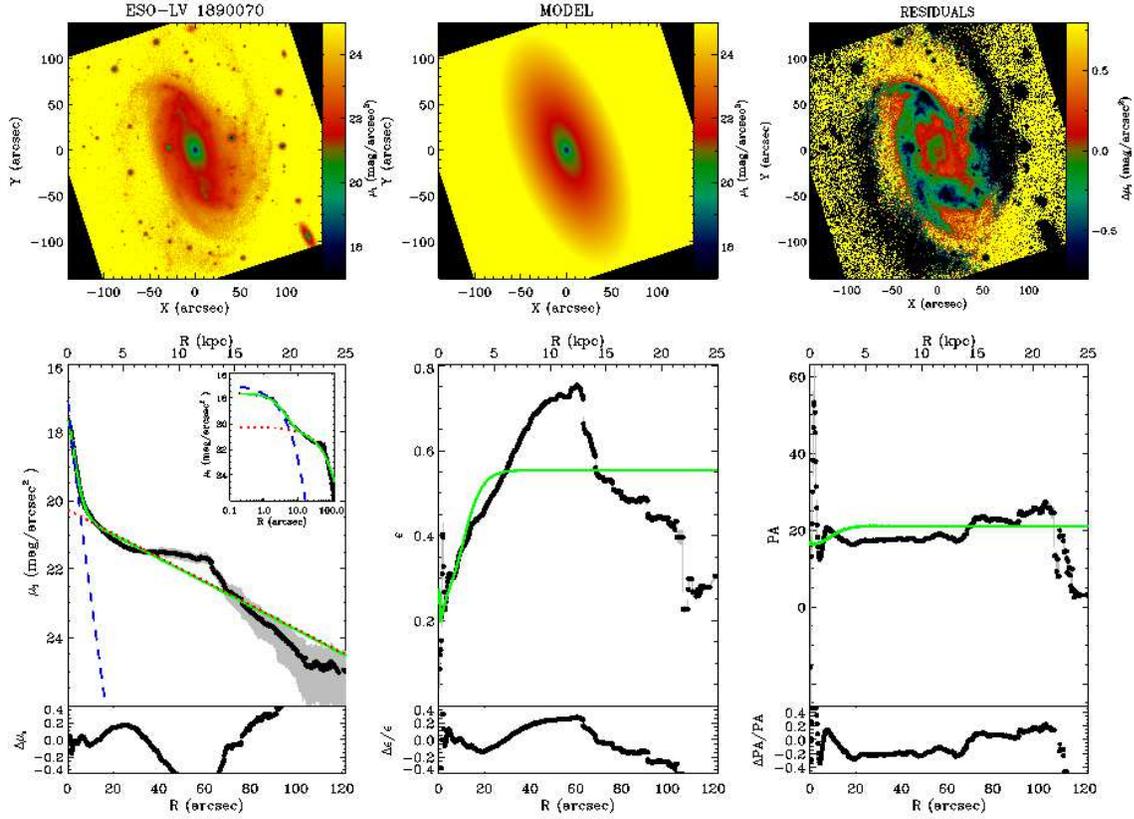}\\
\caption{Photometric decomposition of the sample galaxies. Top left
  panel: galaxy image with North up and East left. Top middle panel:
  best-fitting model of the galaxy image obtained by summing a bulge,
  a disk, and a bar component. Top right: residual image obtained by
  subtracting the best-fitting model from the galaxy image. Bottom
  left panel: ellipse-averaged radial profile of surface brightness of
  the galaxy (black dots) and best-fitting model (green solid line). The
  light contributions of the bulge (dashed blue line), disk (dotted
  red line), and bar (dotted-dashed purple line) are shown. The upper
  inset plots a zoom of the surface-brightness data and fit with a
  logarithmic scale for the distance to the center of the galaxy.
  Bottom middle panel: ellipse-averaged radial profile of ellipticity
  of the galaxy (black dots) and best-fitting model (green solid
  line). Bottom right panel: ellipse-averaged radial profile of
  position angle of the galaxy (black dots) and best-fitting model
  (green solid line). }
\label{fig:gasp2d}
\end{figure*}
\begin{figure*}
\addtocounter{figure}{-1}
\centering
   \includegraphics[angle=0,width=0.9\textwidth]{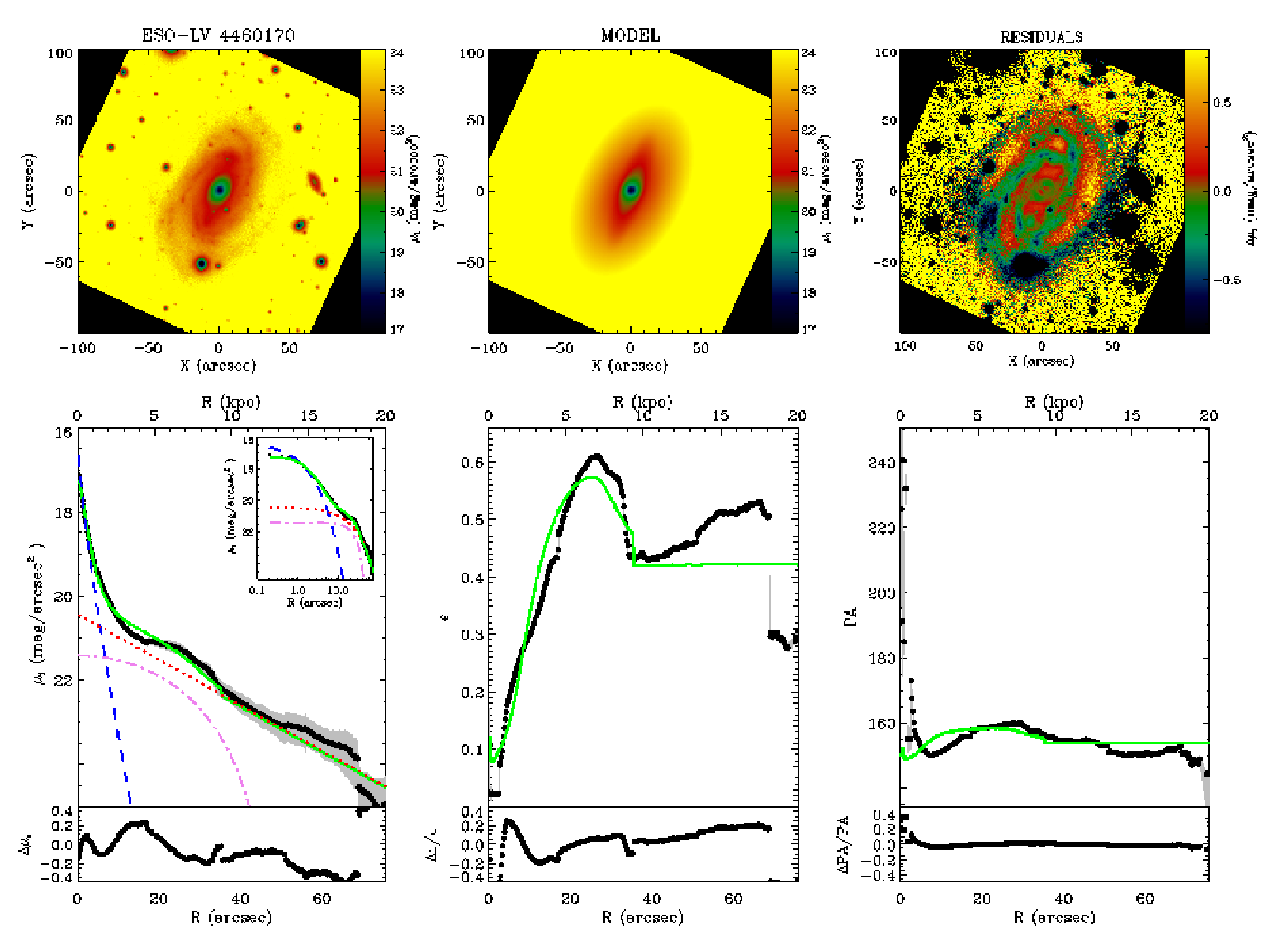}
   \includegraphics[angle=0,width=0.9\textwidth]{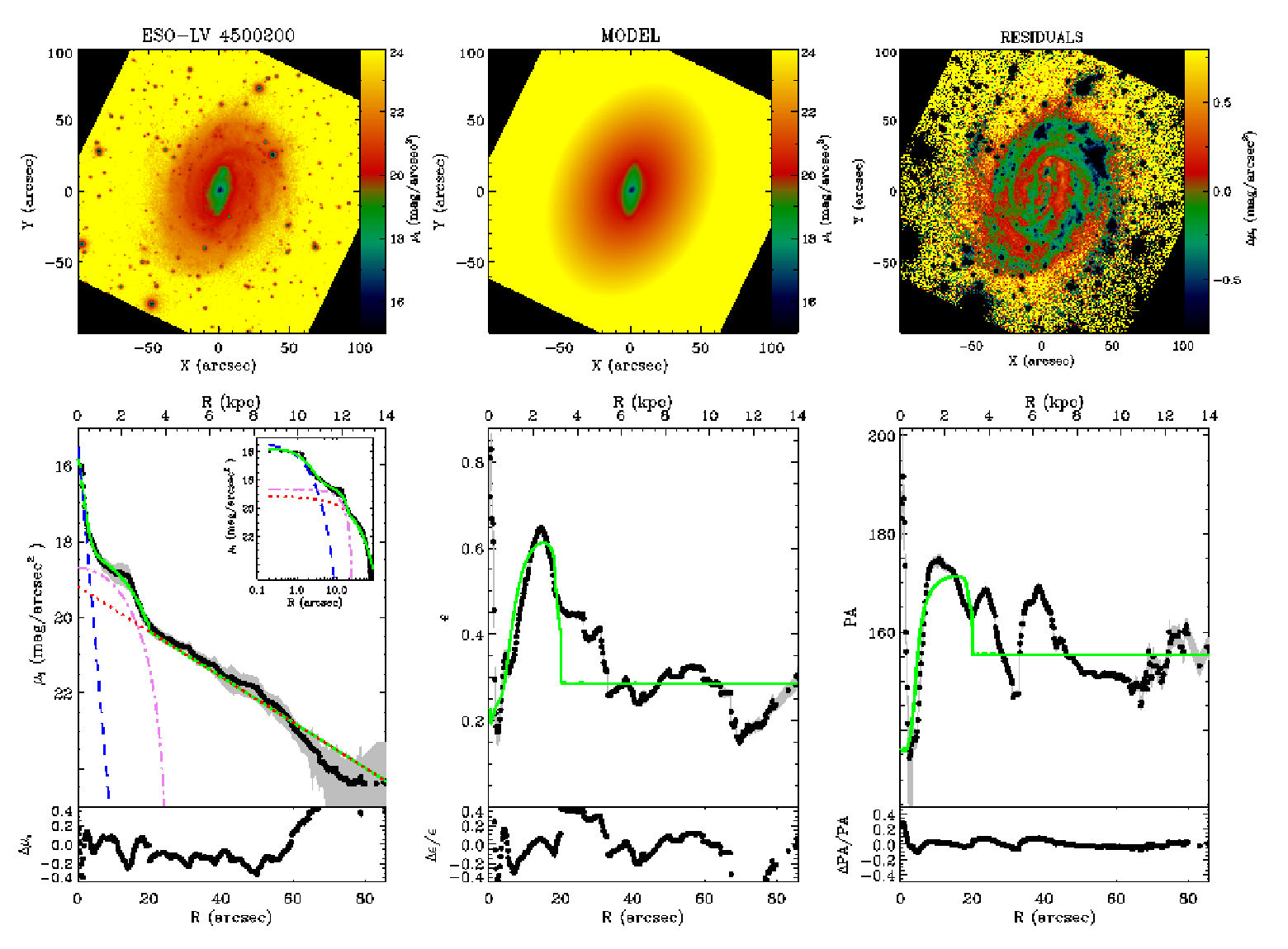}\\
   \caption{ Continued.}
\label{fig:gasp2d}
\end{figure*}

\begin{figure*}
\addtocounter{figure}{-1}
\centering
   \includegraphics[angle=0,width=0.9\textwidth]{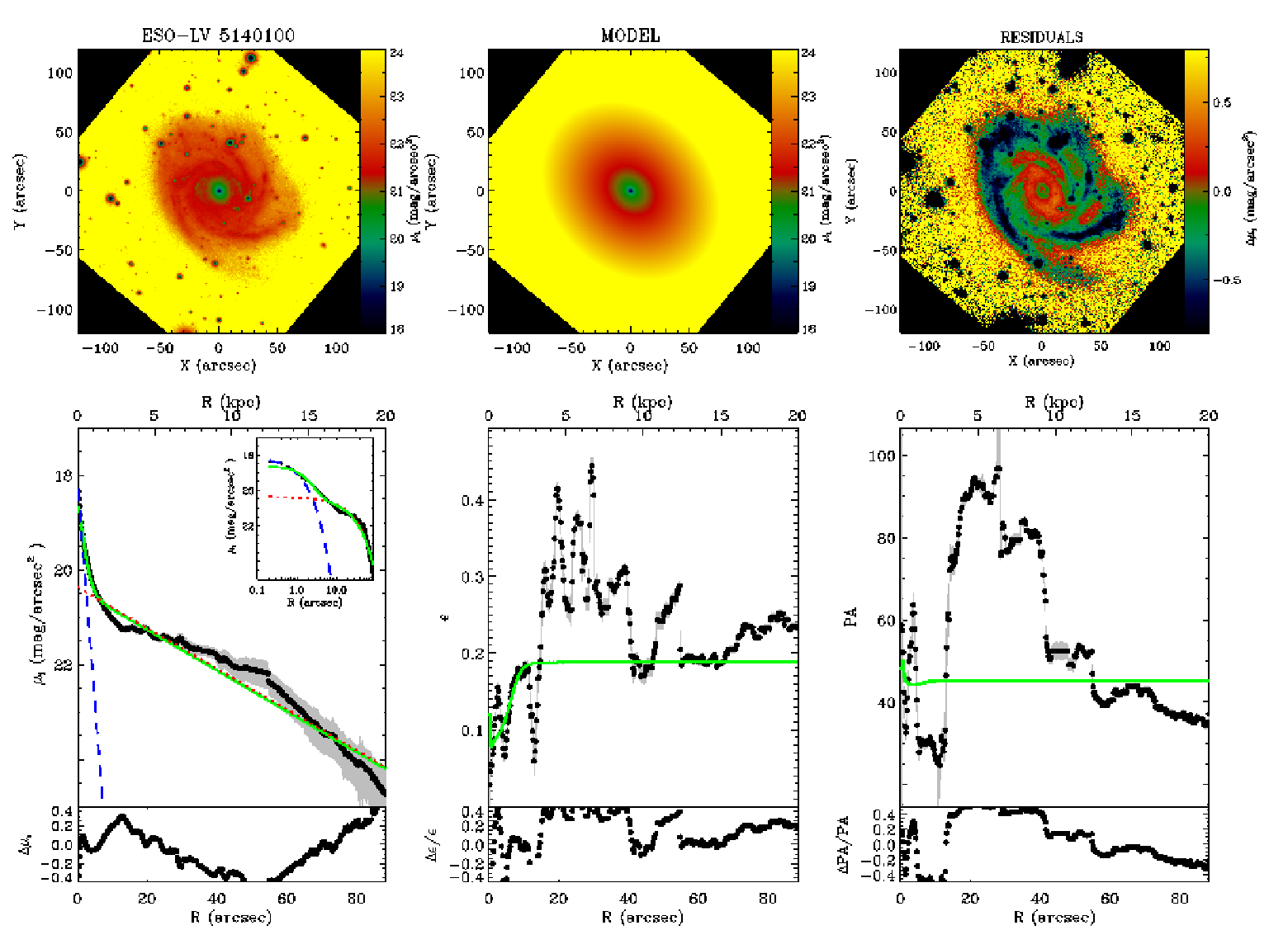}
   \caption{ Continued.}
\label{fig:gasp2d}
\end{figure*}

\section{Measurement of the line-strength indices}
\label{sec:measurements}

Major- and minor-axis spectra were obtained for each sample galaxy by
\citet{pizzetal08}. All the details about the acquisition and
reduction of the galaxy spectra are available in \citet{pizzetal08}
and \citet{moreetal08}.

Mg, Fe, and \Hb\ line-strength indices as defined by
\citet{faberetal85} and \citet{wortetal94} were measured from the flux
calibrated spectra of the four sample galaxies following
\citet{moreetal04, moreetal07}. The average iron index
$\rm{\left<Fe\right> = (Fe5270 + Fe5335)/2}$ \citep{gorgetal90} and
the combined magnesium-iron index $[{\rm MgFe}]^{\prime}=\sqrt{{\rm
    Mg}\,b\,(0.72\times {\rm Fe5270} + 0.28\times{\rm Fe5335})}$
\citep{thmabe03} were measured too.

The difference between the spectral resolution of the galaxy spectra
and the Lick/IDS system ($\rm FWHM = 8.4$ \AA; \citealt{worott97})
was taken into account by degrading our spectra through a Gaussian
convolution to match the Lick/IDS resolution before measuring the
line-strength indices. No focus correction was applied because the
atmospheric seeing was the dominant effect during observations
\citep[see][for details]{mehletal98}. The errors on the line-strength
indices were derived from photon statistics and CCD read-out noise,
and calibrated by means of Monte Carlo simulations.

\begin{figure}
\centering
\includegraphics[angle=0,width=0.46\textwidth]{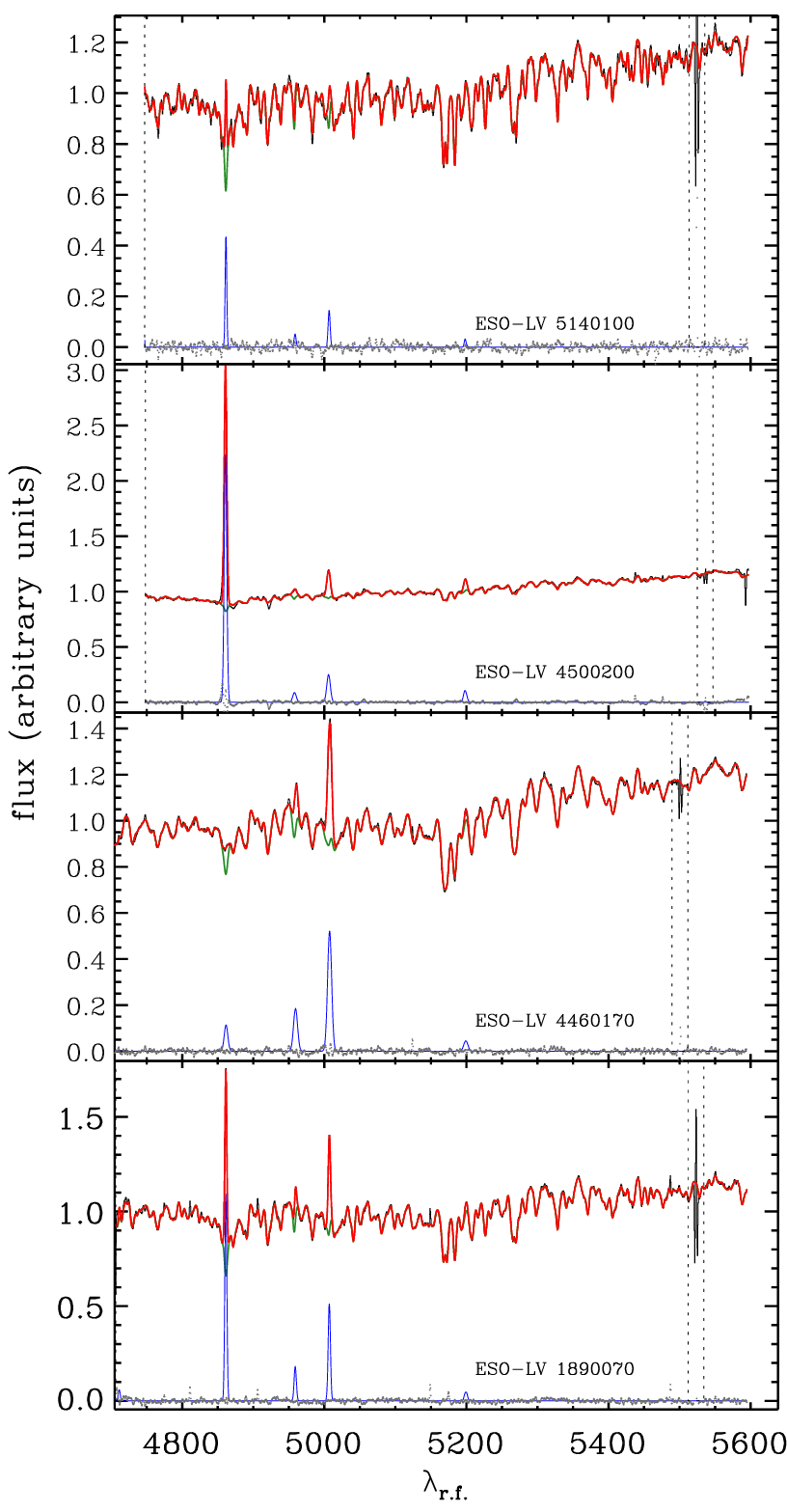}
\caption{ Example of central spectra. Relative fluxes have false zero
  points for viewing convenience.  In each panel the best-fitting
  model (red line) is the sum of the spectra of the ionized-gas (blue
  line) and stellar component (green line). The latter is obtained
  convolving the synthetic templates with the best-fitting LOSVD
  and multiplying them by the best-fitting Legendre polynomials. The
  residuals (grey dots) are obtained by subtracting the model from the
  spectrum. The vertical dashed line corresponds to region masked in
  fitting the spectra.
\label{fig:example_spectra}}
\end{figure}

The contamination of the \Hb\ line-strength index by the \Hb\ emission
line due to the ionized gas present in the galaxy is a problem when
deriving the properties of the stellar populations. Indeed, if the
\Hb\ emission fills the absorption line and a proper separation of
both contributions is not performed before the analysis, the measured
ages result to be artificially older.  To address this issue we
adopted the code Gas AND Absorption Line Fitting (GANDALF) to fit the
galaxy spectra with synthetic population models as done by
\citet{sarzetal06} and \citet{moreetal08}.  The models were built with
different templates from the MILES stellar library by
\citet{vazdekis2010}.  For each spectrum, we fitted a linear
combination of stellar population synthesis models to the
observed galaxy spectrum by performing a $\chi^2$ minimization in
pixel space (Figure \ref{fig:example_spectra}). We adopted the Salpeter initial mass function
\citep{salp55}, ages ranging from 1 to 15 Gyr, and metallicities from
-1.31 to 0.22 dex. We simultaneously fitted the observed spectra using
emission lines in addition to the stellar templates.  Only
\Hb\ emission lines detected with a $S/N > 3$ were subtracted from the
observed spectra.  To calibrate our measurements to the Lick/IDS
system, the values of the line-strength indices measured for a sample
of templates were compared to those obtained by \citet{wortetal94} as
done in \citet{morelli2012}.  The offsets obtained were negligible
when compared to the mean error of the differences between the
tabulated and measured Lick indices. Therefore, no offset correction
was applied to our line-strength measurements.

\begin{figure*}
\centering
\includegraphics[angle=90.0,width=0.40\textwidth]{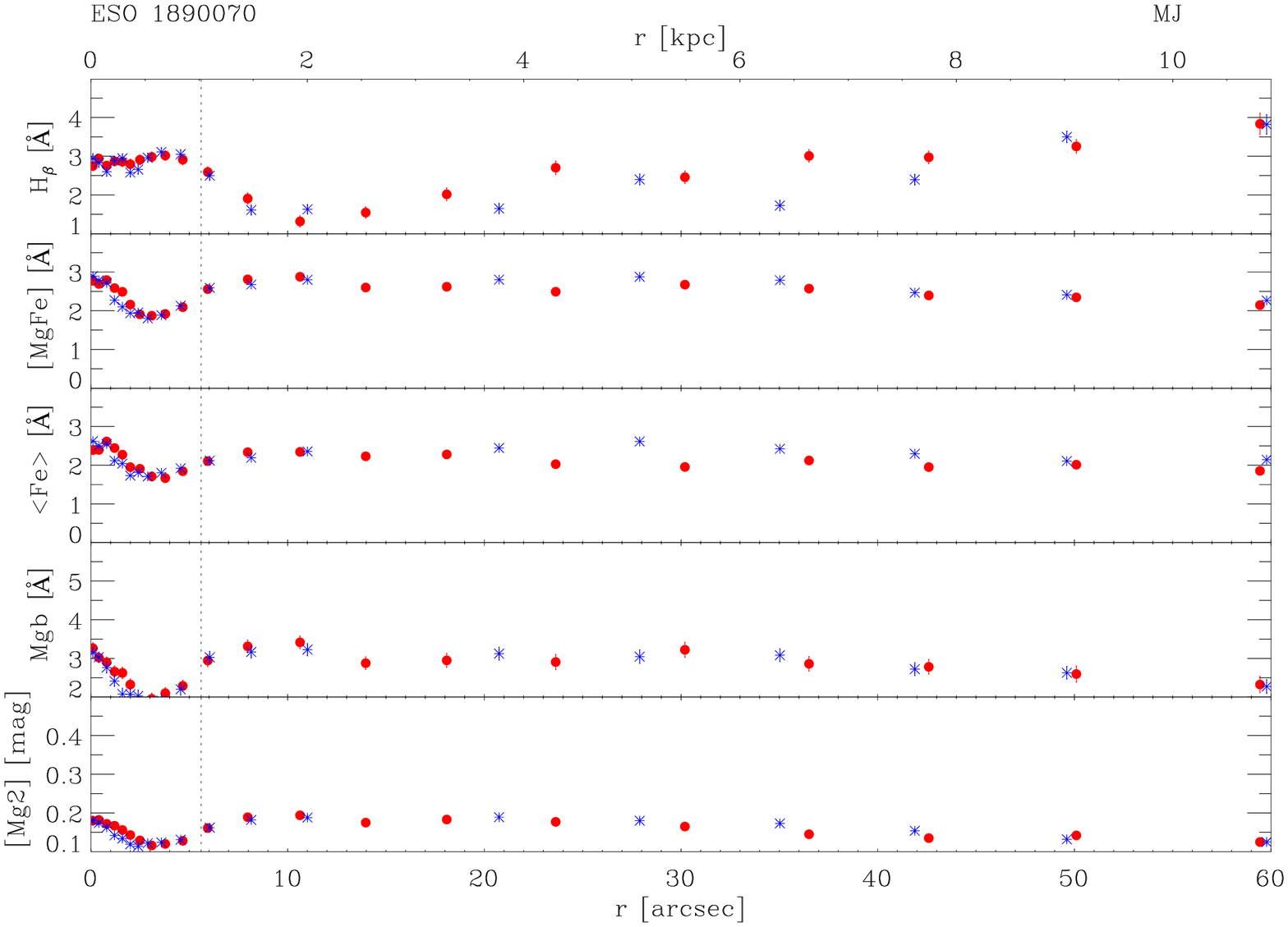}
\includegraphics[angle=90.0,width=0.40\textwidth]{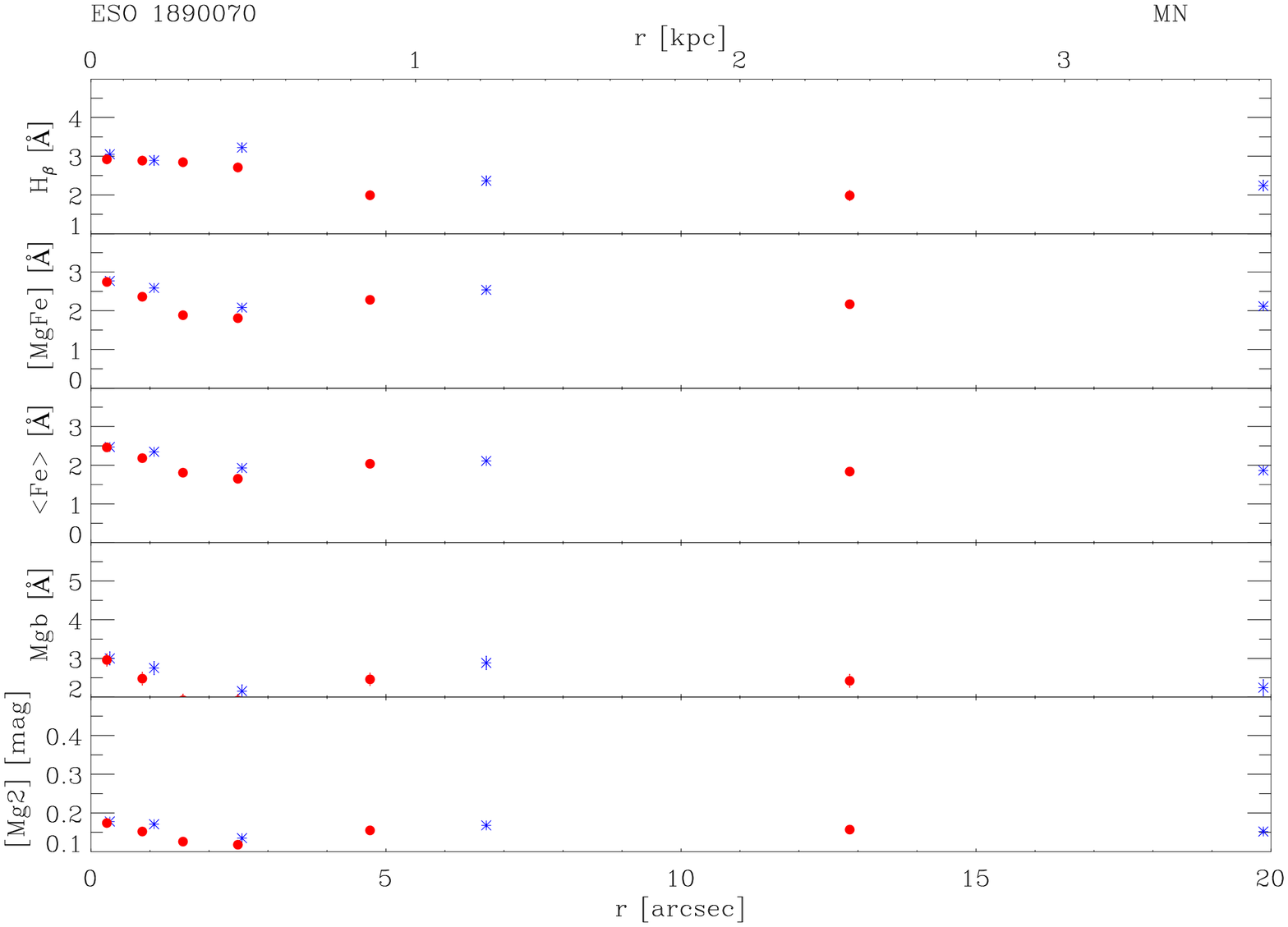}\\
\includegraphics[angle=90.0,width=0.40\textwidth]{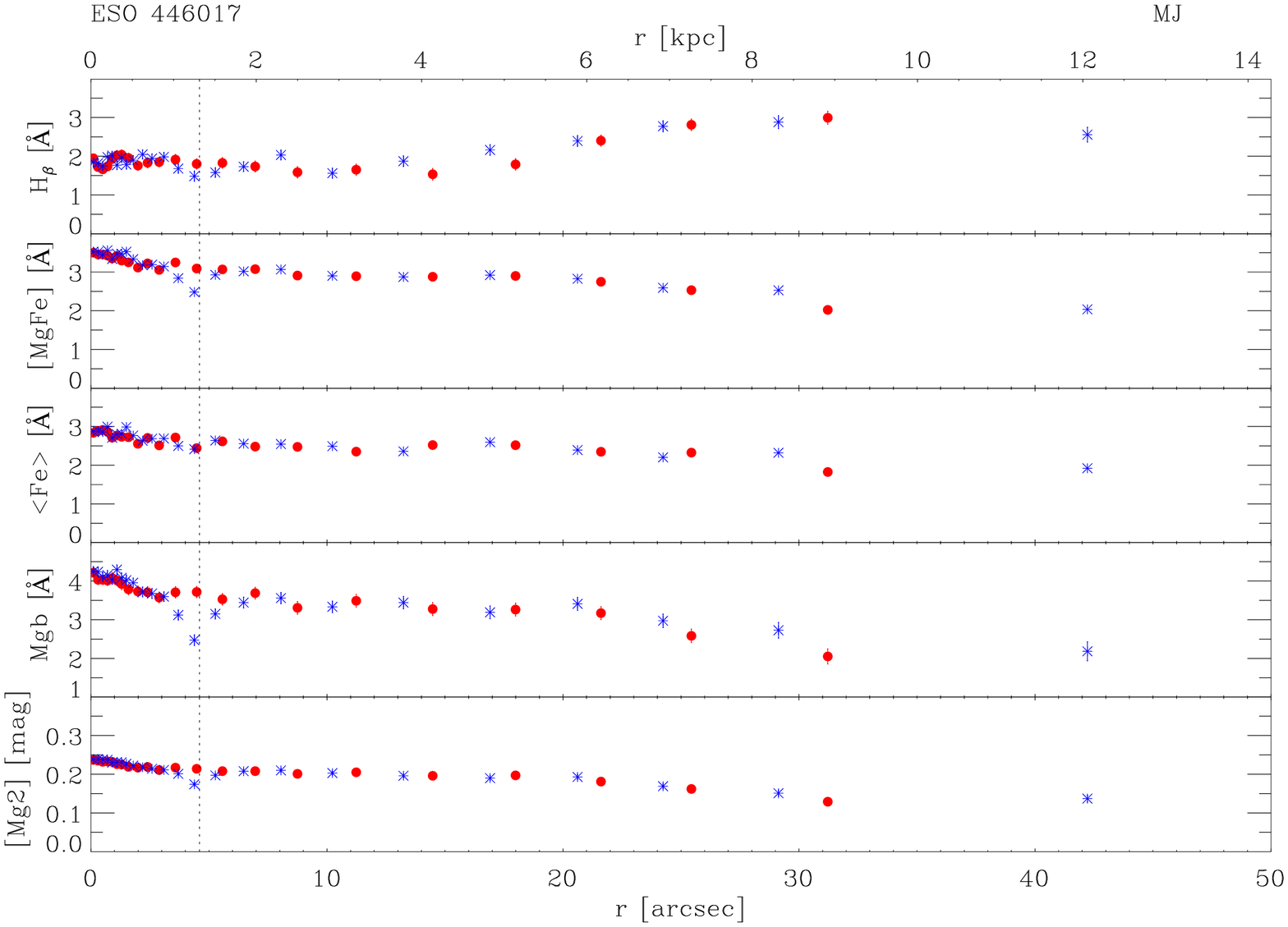}
\includegraphics[angle=90.0,width=0.40\textwidth]{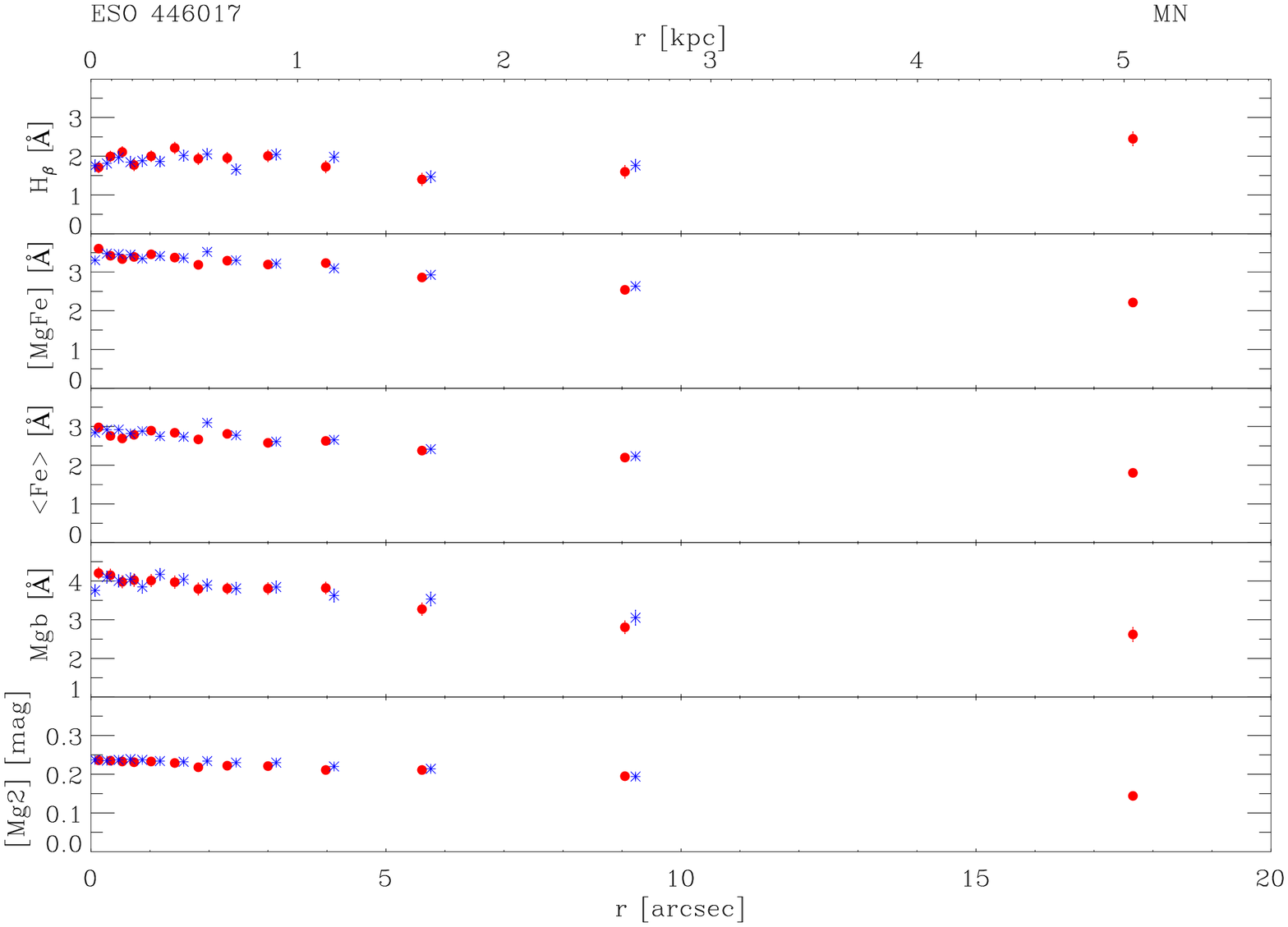}\\
\includegraphics[angle=90.0,width=0.40\textwidth]{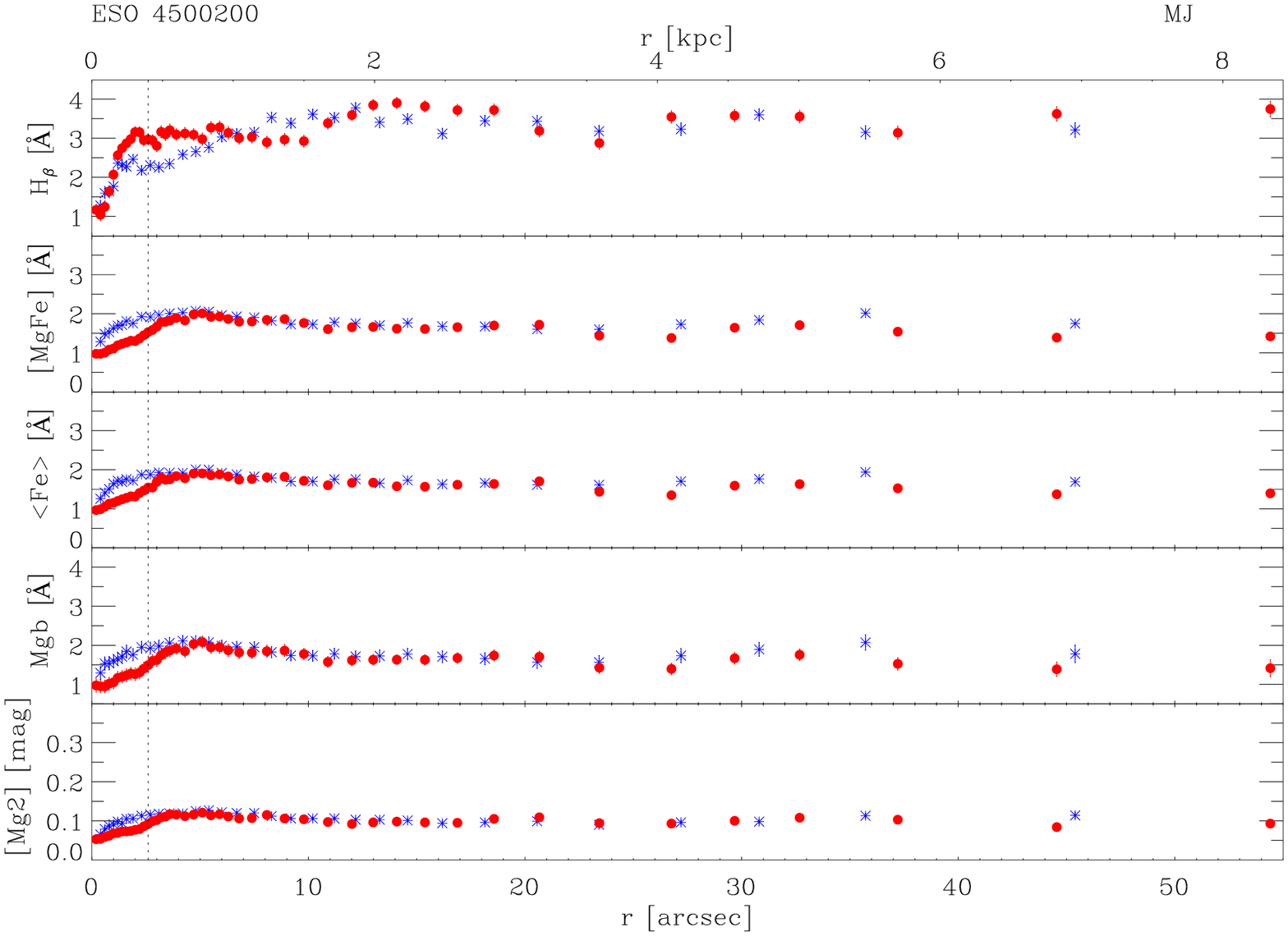}
\includegraphics[angle=90.0,width=0.40\textwidth]{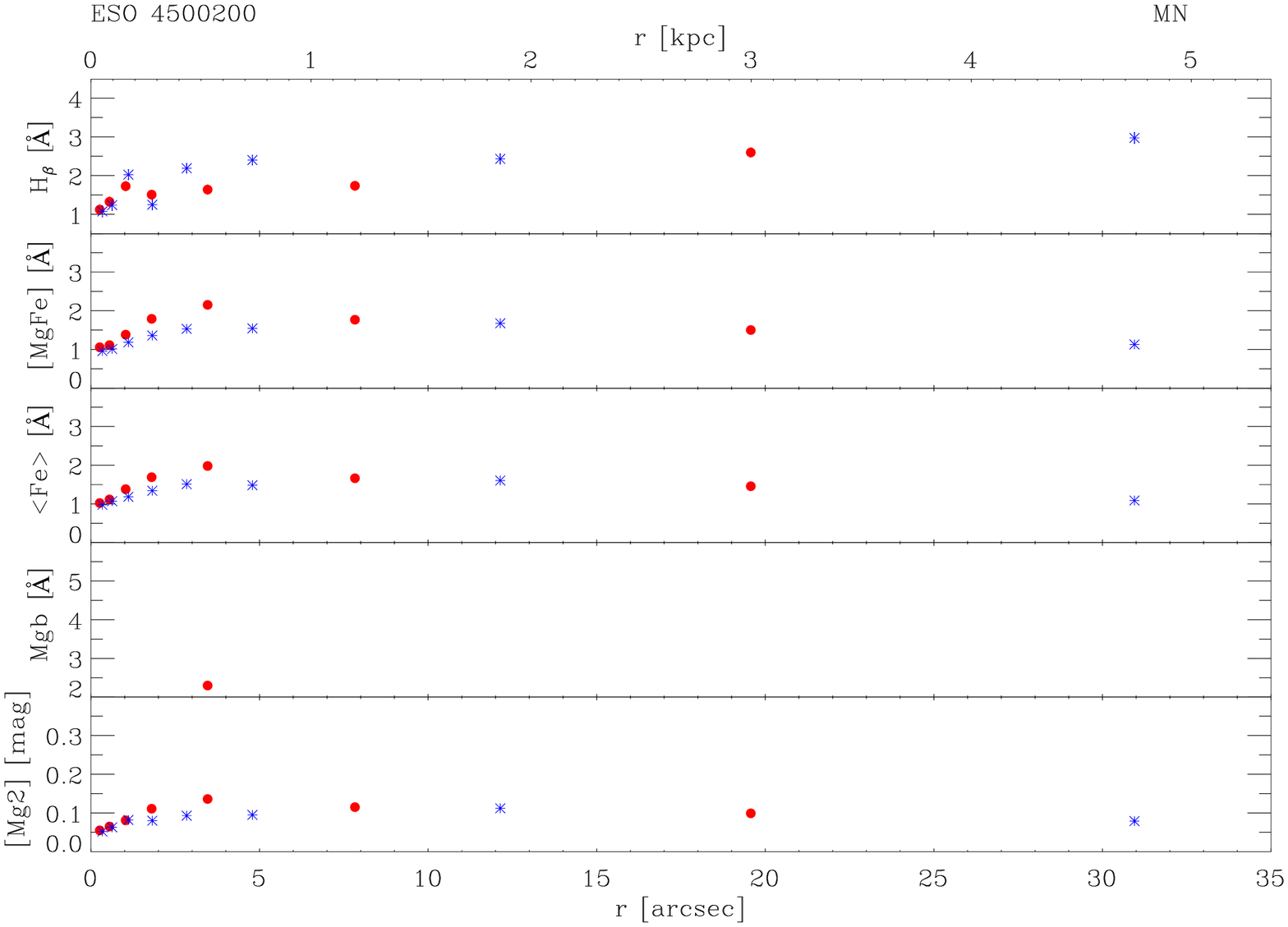}\\
\includegraphics[angle=90.0,width=0.40\textwidth]{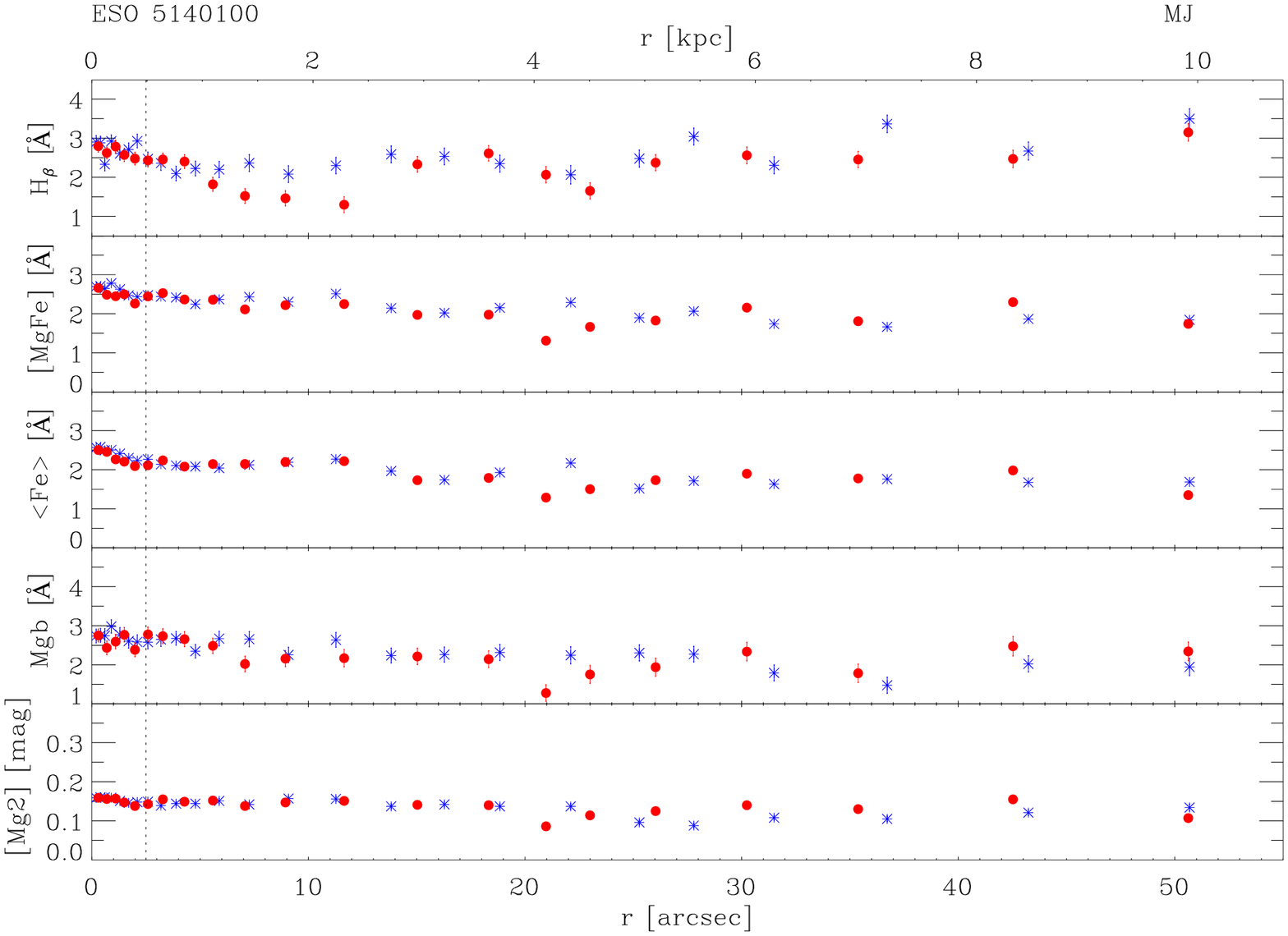}
\includegraphics[angle=90.0,width=0.40\textwidth]{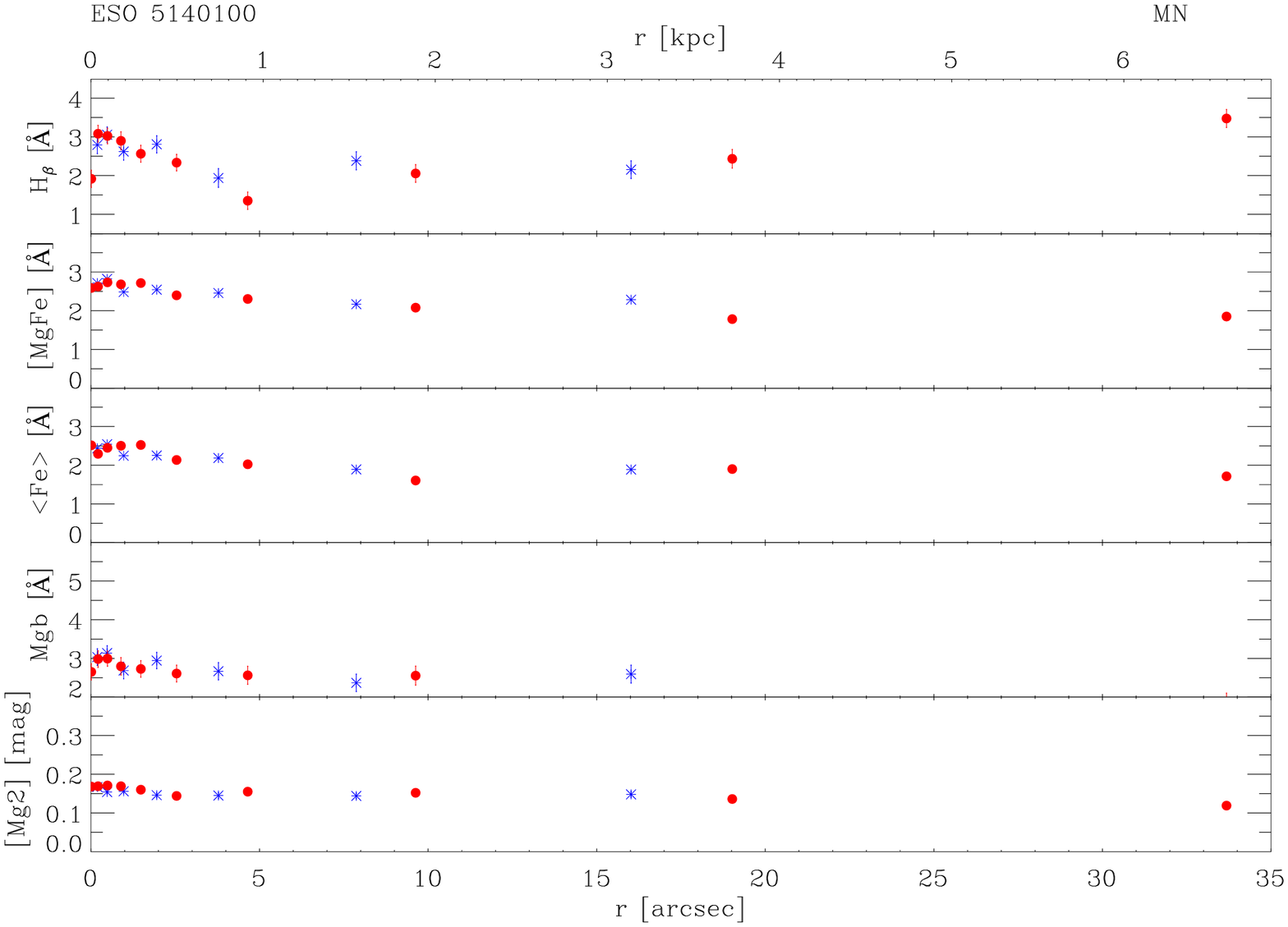}\\
\caption{Line-strength indices measured along the available axes of
  the sample galaxies. For each axis, the curves are folded around the
  nucleus. Blue asterisks and red dots refer to data measured along
  the approaching and receding sides of the galaxy, respectively. The
  radial profiles of the line-strength indices \Hb, \MgFe, \Fe, \Mgb,
  and \Mgd\/ are shown (panels from top to bottom). The vertical
  dotted line corresponds to the radius \Rbd\/ \citep{moreetal08}
  where the surface-brightness is dominated by the light of the
  bulge. For each data set, the name of the galaxy and the location
of the slit position (MJ = major axis and MN = minor axis) are given.}
\label{fig:indices}
\end{figure*}

The measured values of \Hb , \MgFe, \Fe, \Mgb, and \Mgd\ for all the
sample galaxies are plotted in Fig.~\ref{fig:indices}.
The line-strength indices for all the sample galaxies are given in
Table~\ref{tab:val_ind}.


\begin{table*}
\caption{Line-strength indices of the sample galaxies.  }
\begin{tabular}{rrrrrrrr}
\hline
\noalign{\smallskip}
\multicolumn{1}{r}{r~~~~~~~} &
\multicolumn{1}{c}{\Hb}&
\multicolumn{1}{c}{\Mgd} &
\multicolumn{1}{c}{\Mgb} &
\multicolumn{1}{c}{{\rm Fe5270}} &
\multicolumn{1}{c}{{\rm Fe5335}}  \\
\multicolumn{1}{r}{[arcsec] } &
\multicolumn{1}{c}{(\AA)} &
\multicolumn{1}{c}{(mag)} &
\multicolumn{1}{c}{(\AA)} &
\multicolumn{1}{c}{(\AA)} &
\multicolumn{1}{c}{(\AA)} \\
\noalign{\smallskip}
\hline
\noalign{\smallskip}
{\bf ESO-LV~1890070} \\
\noalign{\smallskip}
\hline
\noalign{\smallskip}
$-12.96 $ & $ 1.985  \pm 0.128    $ & $ 0.157    \pm  0.005     $ & $ 2.423   \pm 0.163   $ & $ 2.074     \pm 0.163     $ & $ 1.598   \pm 0.189  $ \\ 
$ -4.83 $ & $ 1.992  \pm 0.116    $ & $ 0.155    \pm  0.004     $ & $ 2.460   \pm 0.156   $ & $ 2.218     \pm 0.147     $ & $ 1.856   \pm 0.172  $ \\ 
$ -2.59 $ & $ 2.710  \pm 0.111    $ & $ 0.118    \pm  0.005     $ & $ 1.912   \pm 0.155   $ & $ 1.785     \pm 0.149     $ & $ 1.516   \pm 0.166  $ \\ 
$ -1.66 $ & $ 2.846  \pm 0.113    $ & $ 0.126    \pm  0.004     $ & $ 1.921   \pm 0.154   $ & $ 1.901     \pm 0.155     $ & $ 1.713   \pm 0.160  $ \\ 
$ ...   $ & $  ...                $ & $ ...                     $ & $ ...                 $ & $ ...                     $ & $ ...                $ \\

\noalign{\smallskip}                                                                                                        
\hline
\noalign{\smallskip}
\hline
\label{tab:val_ind}
\end{tabular}
\end{table*}

\section{Properties of the stellar populations}
\label{sec:populations}

\subsection{Central values of age, metallicity, and ${\bf \alpha}$/Fe ratio}
\label{sec:central}

The central values of velocity dispersion $\sigma$ and line-strength
indices \Mgb, \Mgd, \Hb, \Fe, and \MgFe\/ were derived from the major-
and minor-axis radial profiles as done in
\citet{moreetal08,morelli2012}. 
The data points inside an aperture of radius $0.3~r_{\rm e}$ were
averaged adopting a relative weight proportional to their $S/N$. The
resulting values are listed in Table~\ref{tab:centval_lickind}.

\begin{figure}
\centering
\includegraphics[angle=90,width=0.48\textwidth]{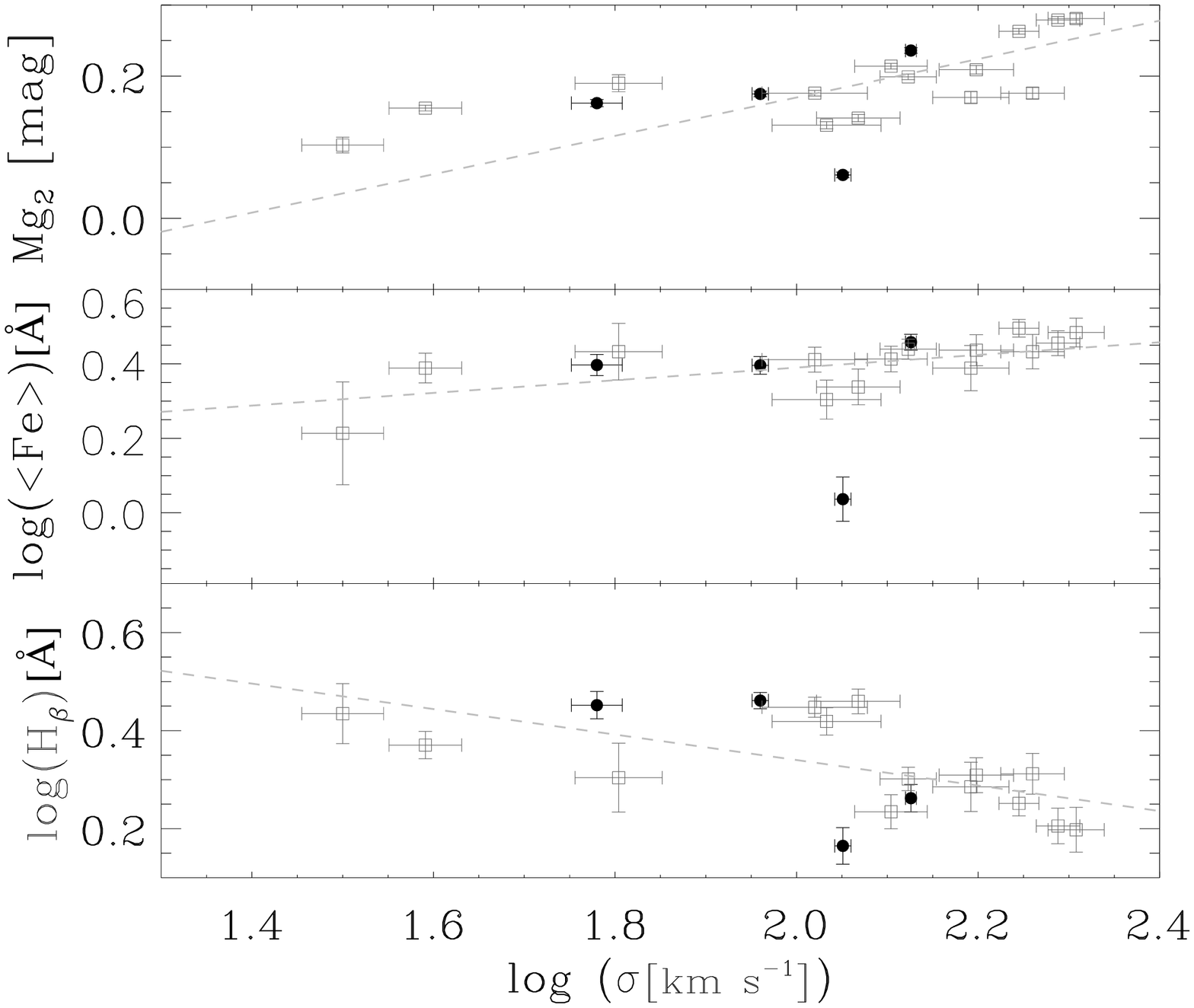}
\caption[The central value correlation \Mgd-$\sigma$ and
  \Fe-$\sigma$]{Central values of the line-strength indices
  \Mgd\ (upper panel), \Fe\ (central panel), and \Hb\ (lower panel) as
  a function of the central velocity dispersion for the sample
  galaxies (filled black circles). Data for bulges in high surface
  brightness disks (open grey squares) and correlations (dashed lines)
  by \citet{moreetal08} are reported for comparison.
\label{fig:censmg2fehb}}
\end{figure}

Fig.~\ref{fig:censmg2fehb} shows the central values of \Mgd, \Hb, and
\Fe\ as a function of the velocity dispersion for the sample
galaxies. The values and correlations for the bulges of the sample of
spiral galaxies with high surface-brightness disks by
\citet{moreetal08} are shown for comparison. The galaxy sample by
\citet{moreetal08} has similar properties to those of the galaxies
studied in this paper.
Three galaxies in our sample follow the trends obtained by
\citet{moreetal08} whereas ESO-LV~4500200 has a lower abundance of
$\alpha-$ and iron elements with respect to galaxies with similar
central velocity dispersion, as shown by the lower values of
\Mgd\ and \Fe.
ESO-LV~4500200 is also characterized by a lower value of \Hb.  It is
worth noticing that \citet{pizzetal08} found a kinematically-decoupled
and dynamically cold component in the nucleus ($r\simeq$ 2 arcsec) of
this galaxy. Such a structure compromises the central stellar velocity
dispersion and affects the values of the Lick indices. This is
indirectly confirmed from the outlier position of ESO-LV~4500200 in
the relations of Fig. \ref{fig:censmg2fehb}

\begin{table*}
\caption{Central values of the velocity dispersion, line-strength
  indices, and equivalent width of the \Hb\ emission line} of the
  sample galaxies averaged within $0.3\, r_{rm e}$.  
\begin{center}
\begin{small}
\begin{tabular}{lrcccccc}
\hline
\noalign{\smallskip}
\multicolumn{1}{c}{Galaxy} &
\multicolumn{1}{c}{$\sigma$} &
\multicolumn{1}{c}{\Fe} &
\multicolumn{1}{c}{\MgFe} &
\multicolumn{1}{c}{\Mgd} &
\multicolumn{1}{c}{\Mgb} &
\multicolumn{1}{c}{\Hb} &
\multicolumn{1}{c}{H$\beta_{\rm em}$ } \\
\multicolumn{1}{c}{ } &
\multicolumn{1}{c}{[\kms]} &
\multicolumn{1}{c}{[\AA]} &
\multicolumn{1}{c}{[\AA]} &
\multicolumn{1}{c}{[mag]} &
\multicolumn{1}{c}{[\AA]} &
\multicolumn{1}{c}{[\AA]} &
\multicolumn{1}{c}{[\AA]} \\
\multicolumn{1}{c}{(1)} &
\multicolumn{1}{c}{(2)} &
\multicolumn{1}{c}{(3)} &
\multicolumn{1}{c}{(4)} &
\multicolumn{1}{c}{(5)} &
\multicolumn{1}{c}{(6)} &
\multicolumn{1}{c}{(7)} &
\multicolumn{1}{c}{(8)} \\
\noalign{\smallskip}
\hline
\noalign{\smallskip}
ESO-LV~1890070 & $ 91\pm2.1$  & $2.49\pm0.14$ & $2.76\pm0.03$ & $0.175\pm0.004$ & $3.00\pm0.15$ & $2.89\pm0.11$ & $3.9$  \\
ESO-LV~4460170 & $133\pm2.2$  & $2.87\pm0.13$ & $3.47\pm0.04$ & $0.234\pm0.004$ & $4.09\pm0.13$ & $1.83\pm0.12$ & $0.56$  \\
ESO-LV~4500200 & $112\pm2.4$  & $1.09\pm0.15$ & $1.08\pm0.01$ & $0.061\pm0.003$ & $1.06\pm0.13$ & $1.46\pm0.13$ & $10.3$  \\
ESO-LV~5140100 & $ 60\pm4.0$  & $2.50\pm0.17$ & $2.70\pm0.04$ & $0.163\pm0.005$ & $2.87\pm0.19$ & $2.83\pm0.19$ & $1.3$  \\
\noalign{\smallskip}
\hline
\noalign{\bigskip}
\end{tabular}
\end{small}
\label{tab:centval_lickind}
\end{center}
\end{table*}

The models by \citet{thmabe03} predict the values of the line-strength
indices for a single stellar population as function of the age,
metallicity, and \aFe ratio.  

\begin{figure*}
\centering
\includegraphics[angle=0.0,width=0.95\textwidth]{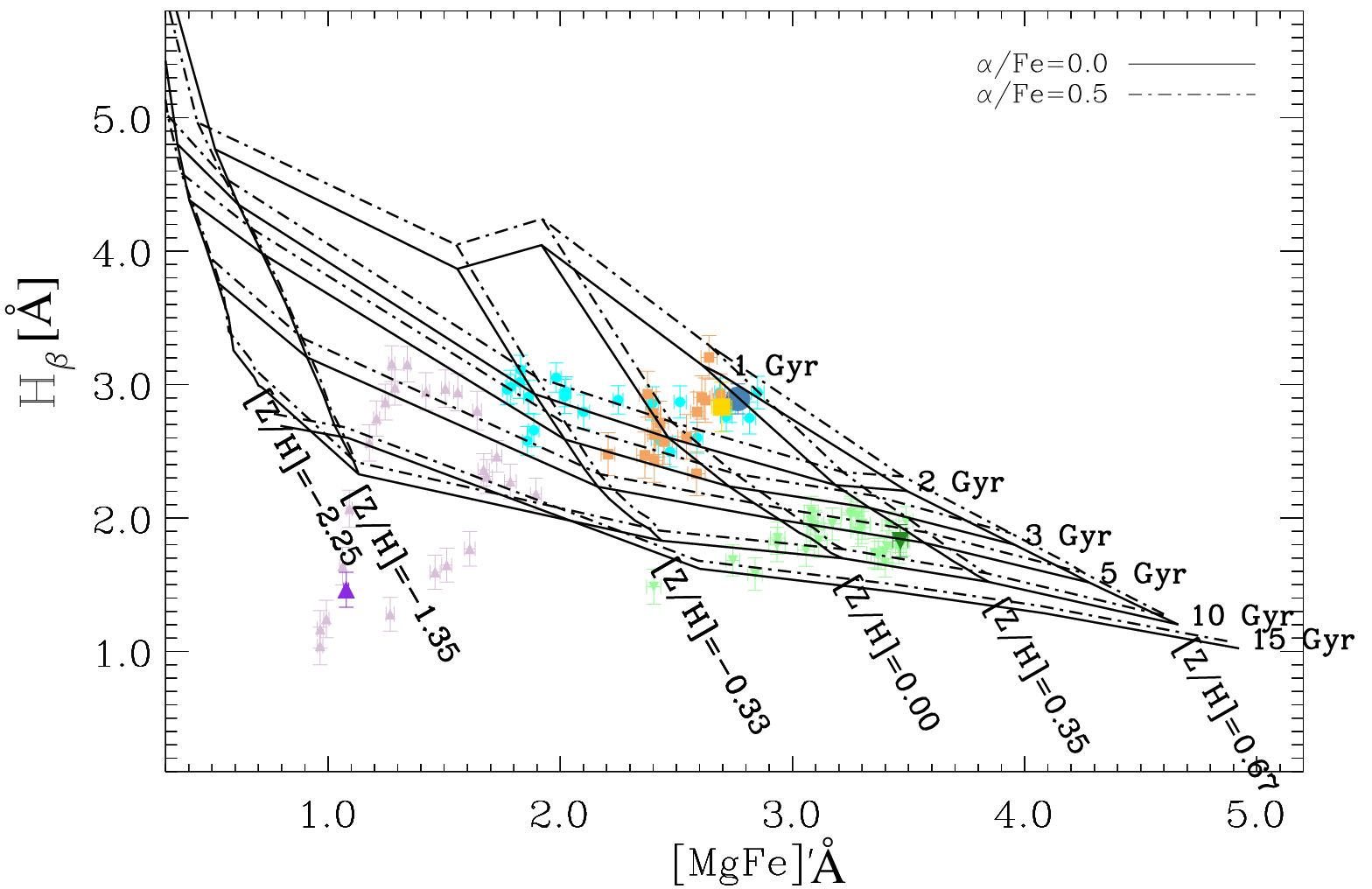}
\includegraphics[angle=0.0,width=0.95\textwidth]{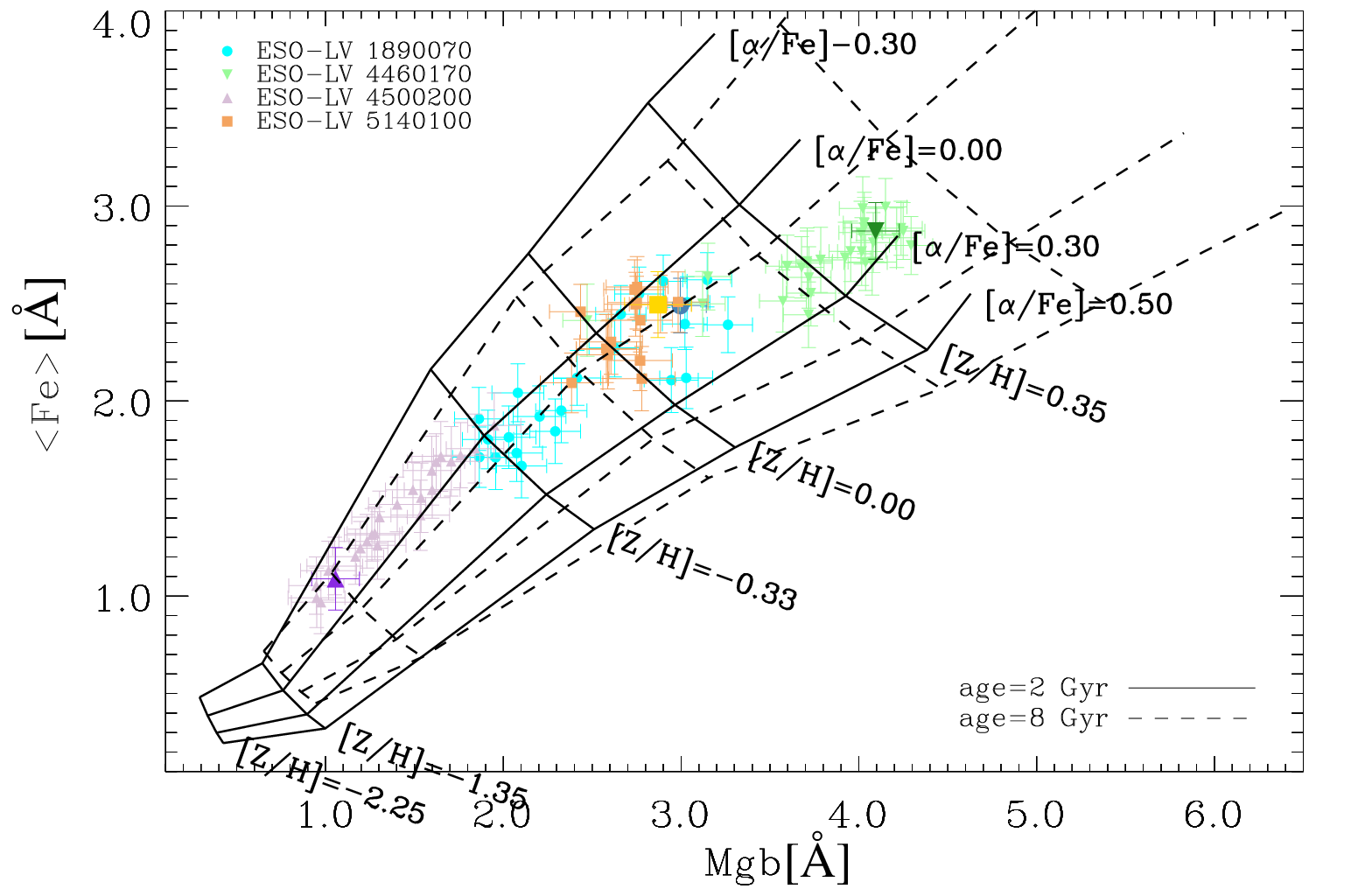}
\caption{The distribution of the values of \Hb\ and \MgFe\ indices (top panel)
  and \Fe\ and \Mgb\ indices (bottom panel) in the bulge
  dominated region (small symbols) and their averaged values inside an
  aperture of radius $0.3\,r_{\rm e}$ (big symbols) for the sample
  galaxies. The lines indicate the models by \cite{thmabe03}. In the
  top panel the age-metallicity grids are plotted with two different
  $\alpha$/Fe enhancements: \aFe$\,=\,0.0$ dex (continuous lines) and
  \aFe$\,=\,0.5$ dex (dashed lines).  In the bottom panel the \aFe\
  ratio-metallicity grids are plotted with two different ages: 2 Gyr
  (continuous lines) and 8 Gyr (dashed lines).
\label{fig:hbemgfemgbcent}}
\end{figure*}

Recent studies indicate the possibility of a multi-SP or extended
  star formation history nature for some bulges. These properties
  could be interpreted using mixed SSPs. However, in this paper we
  decided to follow the SSP approach, since the sample bulges are the
  last ones of a series studied in \citet{moreetal08} and
  \citet{morelli2012}.  This approach allowed us to compare
  the results with the previous data.  
In the top panel of Fig.~\ref{fig:hbemgfemgbcent} the central values
of \Hb\ and \MgFe\ are compared with the model predictions for two
stellar populations with solar (\aFe$\,=\,0$ dex) and super-solar
$\alpha/$Fe ratio (\aFe$\,=\,0.5$ dex), respectively.
In the bottom panel of Fig.~\ref{fig:hbemgfemgbcent} the central
values of \Mgb\ and \Fe\ are compared with the model predictions for
two stellar populations with an intermediate (2 Gyr) and old age (12
Gyr), respectively.

\begin{figure}
\centering
\includegraphics[angle=0,width=0.48\textwidth]{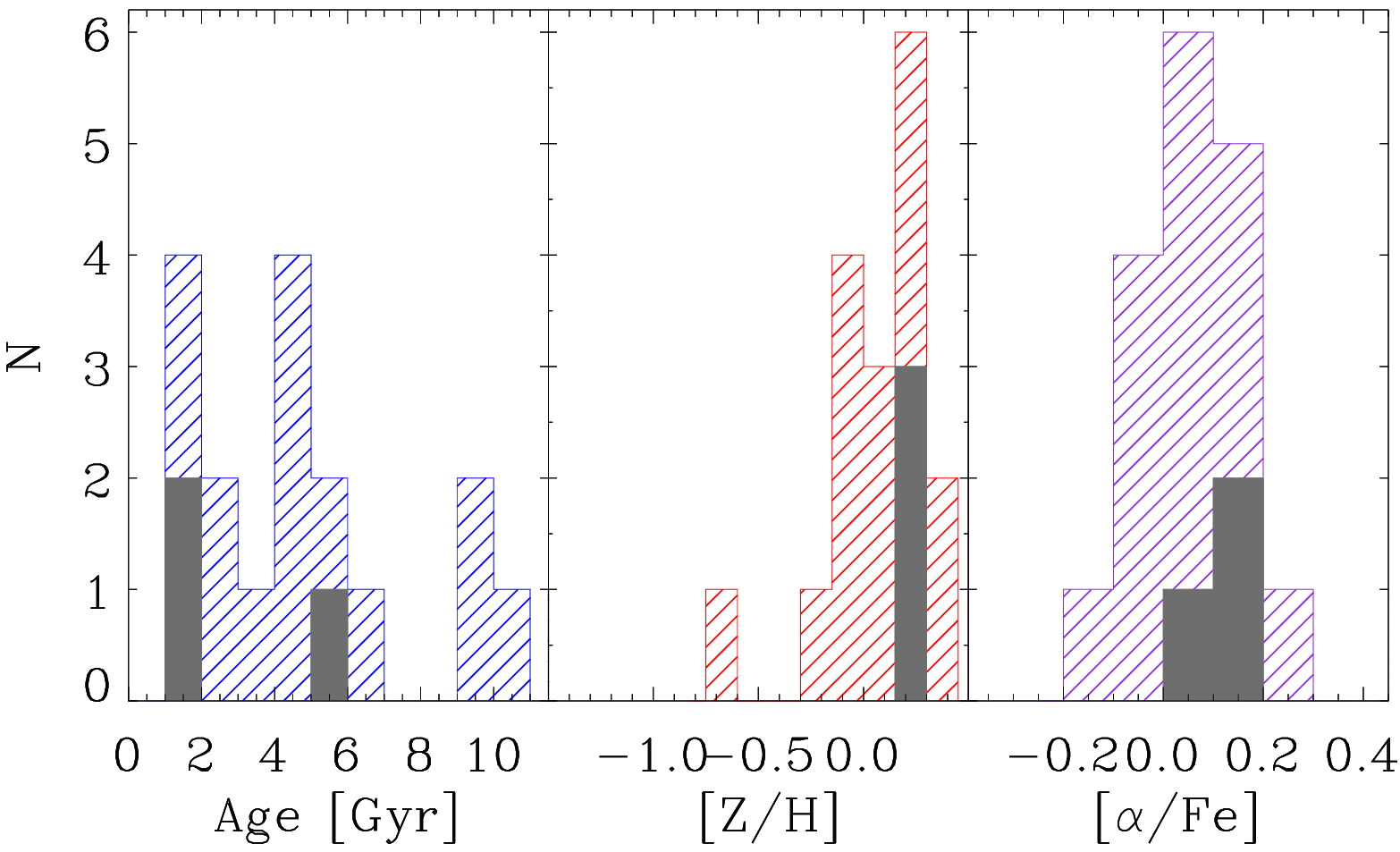}\\
\caption[]{The distribution of the mean age (left-hand panel), total
  metallicity (central panel), and total $\alpha$/Fe enhancement
  (right-hand panel) for the stellar populations of the bulges of the
  sample galaxies (grey histogram) and bulges studied by
  \citet[][hatched histogram]{moreetal08}. The properties of
  ESO-LV~4500200 are not reported in the figure.
\label{fig:hist_ama}}
\end{figure}

The central age, metallicity, and total $\alpha/$Fe ratio of each
bulge were derived by a linear interpolation between the model points
using the iterative procedure described in \citet{mehletal03}. The
derived values and their corresponding errors are listed in
Table~\ref{tab:agemetalfa} and were included in the histograms
  built by \citet{moreetal08} (Fig.~\ref{fig:hist_ama}). 

The values of age, metallicity, and \aFe ratio of ESO-LV~1890070,
ESO-LV~4460170, and ESO-LV~5140100 are consistent with the
distributions obtained by \citet{moreetal08}. On the contrary, the
values of \Fe\ and \Mgb\ of ESO-LV~4500200 in the correspond to
\aFe$\,\simeq\,-0.25$, which is remarkably lower than the average
$\alpha/$Fe ratio found for similar galaxies (Figure
\ref{fig:hist_ama}). Furthermore, the kinematical and chemical
properties of this galaxy place it in a region of the \Hb-\MgFe\
diagram which is not covered by stellar population models. This does
prevented us to derive the age and metallicity of the bulge of
ESO-LV~4500200 and we did not considered it in the analysis of the
gradients of the stellar population properties.

\renewcommand{\tabcolsep}{5pt}
\begin{table}
\caption{Mean age, total metallicity, and total $\alpha/$Fe
  ratio of the stellar populations of the bulges of the sample
  galaxies.}
\begin{center}
\begin{small}
\begin{tabular}{lrrr}
\hline
\noalign{\smallskip}
\multicolumn{1}{c}{Galaxy} &
\multicolumn{1}{c}{Age} &
\multicolumn{1}{c}{\ZH} &
\multicolumn{1}{c}{\aFe} \\
\noalign{\smallskip}
\multicolumn{1}{c}{} &
\multicolumn{1}{c}{[Gyr]} &
\multicolumn{1}{c}{} &
\multicolumn{1}{c}{} \\
\noalign{\smallskip}
\multicolumn{1}{c}{(1)} &
\multicolumn{1}{c}{(2)} &
\multicolumn{1}{c}{(3)} &
\multicolumn{1}{c}{(4)} \\
\noalign{\smallskip}
\hline
\noalign{\smallskip}  
ESO-LV~1890070 &$1.4\pm0.2$ & $0.29\pm0.05$  & $0.14\pm0.07$\\
ESO-LV~4460170 &$5.8\pm2.0$ & $0.26\pm0.07$  & $0.15\pm0.06$\\
ESO-LV~4500200 &$ -       $ & $ -         $   & $-0.25\pm0.18$\\
ESO-LV~5140100 &$1.6\pm0.2$ & $0.21\pm0.06$  & $0.07\pm0.09$\\
\noalign{\smallskip}
\hline
\noalign{\medskip}
\end{tabular}
\end{small}
\label{tab:agemetalfa}
\end{center}
\end{table}
\renewcommand{\tabcolsep}{6pt}

\subsection{Radial gradients of the age, metallicity, and 
${\bf \alpha}$/Fe ratio}

The values of the \Mgd, \Hb, and \Fe\ line-strength indices were
measured along the bulge major axis at the radius
\Rbd\ \citep{moreetal08} where the surface-brightness contributions of
the bulge and disk are equal (Table~\ref{tab:sample}). The
corresponding ages, metallicities, and $\alpha/$Fe ratio were
derived by using the stellar population models by \citet{thmabe03} as
done for the central values.

An issue in measuring the gradients of age, metallicity and
$\alpha/$Fe ratio in bulge, could be the contamination of their
stellar population by the light coming from the underlying disk or bar
stellar component.  This effect is negligible in the galaxy center but
it could increase going to the outer regions of the bulge, where the
light starts to be dominated by the disk component.  In order to
reduce the impact of disk contamination and extend as much as possible
the region in which deriving gradients, we map them inside $r_{\rm
  bd}$, i.e. the radius where the bulge contributes half of the total
surface brightness. This radius is slightly larger than the effective
radius of the galaxy \citep{moreetal08} . Deriving gradients in the
bulge-dominated region, will not remove completely the contamination
by the disk or bar stellar population but it will assure always a
similar degree of contamination in comparing the gradients of
different galaxies.

The gradients were set as the difference between the values at center
and \Rbd\ and their corresponding errors were calculated through Monte
Carlo simulations taking into account the errors
\citep{mehletal03}. The final gradients of the age, metallicity, and
$\alpha/$Fe ratio and their corresponding errors are listed in
Table~\ref{tab:agemetalfa_grad}.

\renewcommand{\tabcolsep}{5pt}
\begin{table}
\caption{Gradients of age, metallicity, and $\alpha/$Fe ratio of
  the stellar populations of the sample bulges derived from the
  central values and values at the radius \Rbd.}
\begin{center}
\begin{small}
\begin{tabular}{rrrr}
\hline
\noalign{\smallskip}
\multicolumn{1}{c}{Galaxy} &
\multicolumn{1}{c}{$\Delta$(Age)} &
\multicolumn{1}{c}{$\Delta$(\ZH)} &
\multicolumn{1}{c}{$\Delta$(\aFe)}\\
\noalign{\smallskip}
\multicolumn{1}{c}{} &
\multicolumn{1}{c}{[Gyr]} &
\multicolumn{1}{c}{} &
\multicolumn{1}{c}{} \\
\noalign{\smallskip}
\multicolumn{1}{c}{(1)} &
\multicolumn{1}{c}{(2)} &
\multicolumn{1}{c}{(3)} &
\multicolumn{1}{c}{(4)} \\
\noalign{\smallskip}
\hline
\noalign{\smallskip} 
ESO-LV~1890070  & $ 0.3\pm0.4$ & $-0.51\pm0.08$ & $-0.07\pm0.10$ \\
ESO-LV~4460170  & $ 1.6\pm3.3$ & $-0.30\pm0.09$ & $-0.08\pm0.11$ \\
ESO-LV~5140100  & $ 0.1\pm0.7$ & $-0.03\pm0.12$ & $ 0.00\pm0.13$ \\
\noalign{\smallskip}
\hline
\noalign{\medskip}
\end{tabular}
\end{small}
\label{tab:agemetalfa_grad}
\end{center}
\end{table}
\renewcommand{\tabcolsep}{6pt}

\begin{figure}
\centering
\includegraphics[angle=0,width=0.48\textwidth]{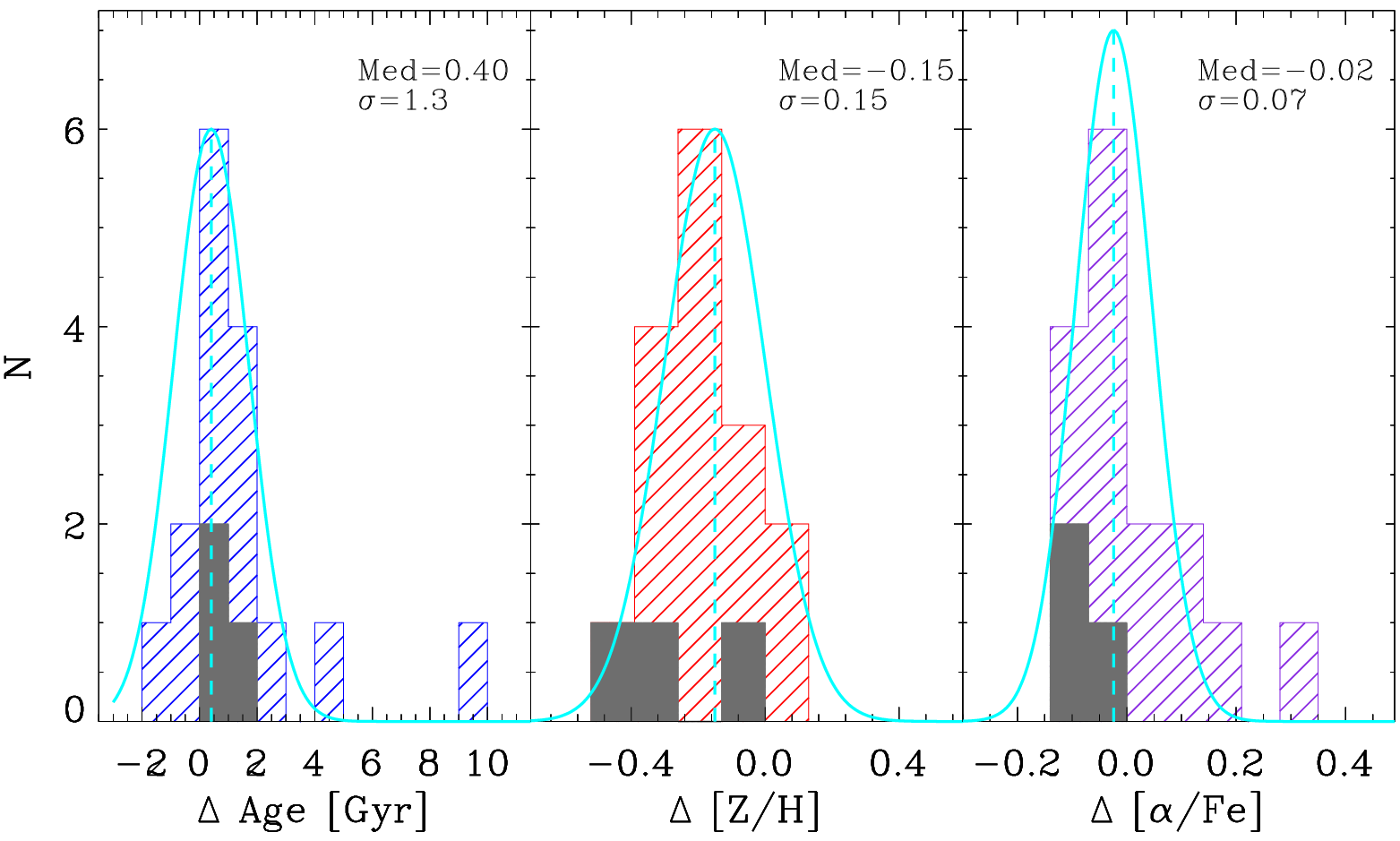}\\
\caption[]{The distribution of the gradients of age (left-hand panel),
  metallicity (central panel) and $\alpha/$Fe ratio (right-hand
  panel) for the bulges of the sample galaxies (grey histogram) and
  bulges studied by \citet[][hatched histogram]{moreetal08}. In each
  panel the dashed line represents the median of the distribution and
  its value is reported. Solid line represents a Gaussian centered in
  the median value of distribution. Its $\sigma$ approximated by the
  value containing the 68\% of the objects of the distribution is
  reported.}
\label{fig:histgrad_ama}
\end{figure}

Shallow gradients in age and no gradient in $\alpha$/Fe ratio
were found for all the sample galaxies, whereas a negative shallow
gradient in metallicity was observed for ESO-LV~1890070,
ESO-LV~4460170, and ESO-LV~5140100. This was somehow expected for the
bulges of spiral galaxies. Indeed, the measured gradients are
consistent with the distributions obtained by \citet{moreetal08}
(Fig.~\ref{fig:histgrad_ama}).

Although we could not derive the age, metallicity, and \aFe, of
ESO-LV 4500200, it is worth noticing that the radial trends of the
line-strength indices are opposite with respect to the other sample
galaxies (Figure \ref{fig:indices}). This is probably due to the presence of a
decoupled component in the center of the galaxy, as suggested by the
kinematical properties discussed in \citet{pizzetal08}.

\section{Conclusions}
\label{sec:conclusions}

The purpose of this paper is to make available to the community the
data and the results of the analysis of the stellar populations of the bulges
hosted in four spiral galaxies with a high surface-brightness disk.

The central values of age, metallicity, and $\alpha/$Fe ratio
were derived for all the sample galaxies. The properties of the
stellar populations of ESO-LV~1890070, ESO-LV~4460170, and
ESO-LV~5140100 are consistent with the previous results obtained for
the bulges of spiral galaxies \citep{jabletal07, gandetal07,
  moreetal08}. 

The gradients of the stellar population properties along the major
axis of all the galaxies were measured in their bulge-dominated
regions. The results are consistent with previous findings for the
bulges of spiral galaxies \citep{jabletal07, moreetal08,
  morelli2012}. The negative metallicity gradient suggests that
dissipative collapse has an important role in the formation of these
galaxies, but the absence of the gradients of $\alpha/$Fe ratio
and age indicates that hierarchical merging might be at
  work during the assembling process of these objects. 

This is also confirmed by the peculiar properties of the bulge of
ESO-LV~4500200 which can hardly be explained without invoking a recent
merging or acquisition \citep{Bercors99, coccetal13, corsini14}.

\acknowledgements          
This work was supported by Padua University through grants
60A02-5052/11, 60A02-4807/12, 60A02-5857/13 and CPDA133894.. LM
acknowledges financial support from Padua University grant CPS0204.
JMA acknowledges support from the European Research Council Starting
Grant (SEDmorph; P.I. V. Wild)


\begin{thebibliography}{}
\bibitem[{{Aguerri} {et~al.}(2009){Aguerri}, {M{\'e}ndez-Abreu}, \&
    {Corsini}}]{aguetal09} 
{Aguerri}, J.~A.~L., {M{\'e}ndez-Abreu}, J., \& {Corsini}, E.~M. 2009,
\aap, 495, 491

\bibitem[{{Annibali} {et~al.}(2007){Annibali}, {Bressan}, {Rampazzo},
  {Zeilinger}, \& {Danese}}]{annietal07}
{Annibali}, F., {Bressan}, A., {Rampazzo}, R., {Zeilinger}, W.~W., \& {Danese},
  L. 2007, \aap, 463, 455

\bibitem[{{Athanassoula} {et~al.}(1990){Athanassoula}, {Morin},
    {Wozniak}, {Puy}, {Pierce}, {Lombard}, \& {Bosma}}]{athetal90}
{Athanassoula}, E., {Morin}, S., {Wozniak}, H., {et~al.} 1990, \mnras,
245, 130

\bibitem[{{Bertola} \& {Corsini}(1999)}]{Bercors99}
{Bertola}, F. \& {Corsini}, E.~M. 1999, in Galaxy Interactions at Low and High
  Redshift, ed. {J.~E.~Barnes \& D.~B.~Sanders}, IAU Symp. 186 (Kuwler, Dordrecht, 149

\bibitem[{{Caon} {et~al.}(1993){Caon}, {Capaccioli}, \&
    {D'Onofrio}}]{caonetal93} 
{Caon}, N., {Capaccioli}, M., \& {D'Onofrio}, M. 1993, \mnras, 265,
1013

\bibitem[{{Coccato} {et~al.}(2013){Coccato}, {Morelli}, {Pizzella}, {Corsini},
  {Buson}, \& {Dalla Bont{\`a}}}]{coccetal13}
{Coccato}, L., {Morelli}, L., {Pizzella}, A., {et~al.} 2013, \aap, 549, A3

\bibitem[Coccato et al.(2011)]{coccato11} Coccato, L., Morelli, 
L., Corsini, E.~M., et al.\ 2011, \mnras, 412, L113 

\bibitem[{{Cole} {et~al.}(2000){Cole}, {Lacey}, {Baugh}, \&
  {Frenk}}]{coletal00}
{Cole}, S., {Lacey}, C.~G., {Baugh}, C.~M., \& {Frenk}, C.~S. 2000, \mnras,
  319, 168

\bibitem[{{Collobert} {et~al.}(2006){Collobert}, {Sarzi}, {Davies},
  {Kuntschner}, \& {Colless}}]{colletal06}
{Collobert}, M., {Sarzi}, M., {Davies}, R.~L., {Kuntschner}, H., \& {Colless},
  M. 2006, \mnras, 370, 1213

\bibitem[{{Corsini}(2014)}]{corsini14}
{Corsini}, E.~M. 2014, in Multi-Spin Galaxies, ed.
  E.~{Iodice} \& E.~M. {Corsini}, ASP Conf. Ser. 486, (ASP, San Francisco), 51

\bibitem[{{Faber} {et~al.}(1985){Faber}, {Friel}, {Burstein}, \&
  {Gaskell}}]{faberetal85}
{Faber}, S.~M., {Friel}, E.~D., {Burstein}, D., \& {Gaskell}, C.~M. 1985,
  \apjs, 57, 711

\bibitem[{{Ferrers}(1877)}]{ferrers77}
{Ferrers}, N.~M. 1877, Quart. J. Pure and Appl. Math, 14, 1

\bibitem[{{Freeman}(1970)}]{freeman70}
{Freeman}, K.~C. 1970, \apj, 160, 811

\bibitem[{{Ganda} {et~al.}(2007){Ganda}, {Peletier}, {McDermid},
  {Falc{\'o}n-Barroso}, {de Zeeuw}, {Bacon}, {Cappellari}, {Davies},
  {Emsellem}, {Krajnovi{\'c}}, {Kuntschner}, {Sarzi}, \& {van de
  Ven}}]{gandetal07}
{Ganda}, K., {Peletier}, R.~F., {McDermid}, R.~M., {et~al.} 2007, \mnras, 380,
  506

\bibitem[{{Gilmore} \& {Wyse}(1998)}]{gilwys98}
{Gilmore}, G. \& {Wyse}, R.~F.~G. 1998, \aj, 116, 748

\bibitem[{{Gorgas} {et~al.}(1990){Gorgas}, {Efstathiou}, \&
  {Salamanca}}]{gorgetal90}
{Gorgas}, J., {Efstathiou}, G., \& {Salamanca}, A.~A. 1990, \mnras, 245, 217

\bibitem[{{Jablonka} {et~al.}(2007){Jablonka}, {Gorgas}, \&
    {Goudfrooij}}]{jabletal07} {Jablonka}, P., {Gorgas}, J., \&
  {Goudfrooij}, P. \ 2007, \aap, 474, 763

\bibitem[{{Kormendy} {et~al.}(2009){Kormendy}, {Fisher}, {Cornell}, \&
  {Bender}}]{kormetal09}
{Kormendy}, J., {Fisher}, D.~B., {Cornell}, M.~E., \& {Bender}, R. 2009, \apjs,
  182, 216

\bibitem[{{Kormendy} \& {Kennicutt}(2004)}]{korken04}
{Kormendy}, J. \& {Kennicutt}, R.~C. 2004, \araa, 42, 603

\bibitem[{{Kuntschner} {et~al.}(2010){Kuntschner}, {Emsellem}, {Bacon},
  {Cappellari}, {Davies}, {de Zeeuw}, {Falc{\'o}n-Barroso}, {Krajnovi{\'c}},
  {McDermid}, {Peletier}, {Sarzi}, {Shapiro}, {van den Bosch}, \& {van de
  Ven}}]{kuntetal10}
{Kuntschner}, H., {Emsellem}, E., {Bacon}, R., {et~al.} 2010, \mnras, 408, 97

\bibitem[McDermid et al.(2006)]{mcdermid06} McDermid, R.~M., 
Emsellem, E., Shapiro, K.~L., et al.\ 2006, \mnras, 373, 906 

\bibitem[{{MacArthur} {et~al.}(2009){MacArthur}, {Gonz{\'a}lez}, \&
  {Courteau}}]{macaetal09}
{MacArthur}, L.~A., {Gonz{\'a}lez}, J.~J., \& {Courteau}, S. 2009, \mnras, 395,
  28

\bibitem[{{Markwardt}(2009)}]{markwardt09}
{Markwardt}, C.~B. 2009, in Astronomical Data Analysis Software and
Systems XVIII, ed.  D.~A. {Bohlender}, D.~{Durand}, \& P.~{Dowler},
ASP Conf. Ser. 411, (ASP, San Francisco), 251

\bibitem[{{Mehlert} {et~al.}(1998){Mehlert}, {Saglia}, {Bender}, \&
  {Wegner}}]{mehletal98}
{Mehlert}, D., {Saglia}, R.~P., {Bender}, R., \& {Wegner}, G. 1998, \aap, 332,
  33

\bibitem[{{Mehlert} {et~al.}(2003){Mehlert}, {Thomas}, {Saglia}, {Bender}, \&
  {Wegner}}]{mehletal03}
{Mehlert}, D., {Thomas}, D., {Saglia}, R.~P., {Bender}, R., \& {Wegner}, G.
  2003, \aap, 407, 423

\bibitem[M{\'e}ndez-Abreu et al.(2008)]{mendetal08} M{\'e}ndez-Abreu, J., Aguerri, J.~A.~L., Corsini, E.~M., \& Simonneau, E.\ 2008, \aap, 478, 353 

\bibitem[Mendez-Abreu et al.(2014)]{mendetal14} Mendez-Abreu, J., Debattista, V.~P., Corsini, E.~M., 
\& Aguerri, J.~A.~L.\ 2014, arXiv:1409.2876 

\bibitem[{{Moorthy} \& {Holtzman}(2006)}]{mooretal06}
{Moorthy}, B.~K. \& {Holtzman}, J.~A. 2006, \mnras, 371, 583

\bibitem[{{Mor\'e} {et~al.}(1980){Mor\'e}, {Garbow}, \& {Hillstrom}}]{moretal80}
{Mor\'e}, J.~J., {Garbow}, B.~S., \& {Hillstrom}, K.~E. 1980, Argonne National
  Laboratory Report ANL-80-74

\bibitem[Morelli et al.(2004)]{morelli04} Morelli, L., Halliday, 
C., Corsini, E.~M., et al.\ 2004, \mnras, 354, 753 

\bibitem[{{Morelli} {et~al.}(2013){Morelli}, {Calvi}, {Masetti}, {Parisi},
  {Landi}, {Maiorano}, {Minniti}, \& {Galaz}}]{moreetal13}
{Morelli}, L., {Calvi}, V., {Masetti}, N., {et~al.} 2013, \aap, 556, A135

\bibitem[{{Morelli} {et~al.}(2012){Morelli}, {Corsini}, {Pizzella}, {Dalla
  Bont{\`a}}, {Coccato}, {M{\'e}ndez-Abreu}, \& {Cesetti}}]{morelli2012}
{Morelli}, L., {Corsini}, E.~M., {Pizzella}, A., {et~al.} 2012, \mnras, 423,
  962

\bibitem[{{Morelli} {et~al.}(2004){Morelli}, {Halliday}, {Corsini}, {Pizzella},
  {Thomas}, {Saglia}, {Davies}, {Bender}, {Birkinshaw}, \&
  {Bertola}}]{moreetal04}
{Morelli}, L., {Halliday}, C., {Corsini}, E.~M., {et~al.} 2004, \mnras, 354,
  753

\bibitem[{{Morelli} {et~al.}(2007){Morelli}, {Pompei}, {Pizzella}, {Coccato},
  {Corsini}, {M{\'e}ndez-Abreu}, {Saglia}, {Sarzi}, \& {Bertola}}]{moreetal07}
{Morelli}, L., {Pompei}, E., {Pizzella}, A., {et~al.} 2007, Nuovo Cimento B
  Serie, 122, 1281

\bibitem[{{Morelli} {et~al.}(2008){Morelli}, {Pompei}, {Pizzella},
  {M{\'e}ndez-Abreu}, {Corsini}, {Coccato}, {Saglia}, {Sarzi}, \&
  {Bertola}}]{moreetal08}
{Morelli}, L., {Pompei}, E., {Pizzella}, A., {et~al.} 2008, \mnras, 389, 341

\bibitem[{{Pizzella} {et~al.}(2008){Pizzella}, {Corsini}, {Sarzi}, {Magorrian},
  {M{\'e}ndez-Abreu}, {Coccato}, {Morelli}, \& {Bertola}}]{pizzetal08}
{Pizzella}, A., {Corsini}, E.~M., {Sarzi}, M., {et~al.} 2008, \mnras, 387, 1099

\bibitem[{{Rampazzo} {et~al.}(2005){Rampazzo}, {Annibali}, {Bressan},
  {Longhetti}, {Padoan}, \& {Zeilinger}}]{rampetal05}
{Rampazzo}, R., {Annibali}, F., {Bressan}, A., {et~al.} 2005, \aap, 433, 497

\bibitem[{{Rawle} {et~al.}(2010){Rawle}, {Smith}, \& {Lucey}}]{rawletal10}
{Rawle}, T.~D., {Smith}, R.~J., \& {Lucey}, J.~R. 2010, \mnras, 401, 852

\bibitem[{{Salpeter}(1955)}]{salp55} {Salpeter}, E.~E. 1955, \apj, 121, 161

\bibitem[{{S{\'a}nchez-Bl{\'a}zquez} {et~al.}(2006){S{\'a}nchez-Bl{\'a}zquez},
  {Gorgas}, {Cardiel}, \& {Gonz{\'a}lez}}]{sancetal06p}
{S{\'a}nchez-Bl{\'a}zquez}, P., {Gorgas}, J., {Cardiel}, N., \& {Gonz{\'a}lez},
  J.~J. 2006, \aap, 457, 809

\bibitem[{{Sarzi} {et~al.}(2006){Sarzi}, {Falc{\'o}n-Barroso}, {Davies},
  {Bacon}, {Bureau}, {Cappellari}, {de Zeeuw}, {Emsellem}, {Fathi},
  {Krajnovi{\'c}}, {Kuntschner}, {McDermid}, \& {Peletier}}]{sarzetal06}
{Sarzi}, M., {Falc{\'o}n-Barroso}, J., {Davies}, R.~L., {et~al.} 2006, \mnras,
  366, 1151

\bibitem[{{S\'ersic}(1968)}]{sersic68} 
{S\'ersic}, J.~L. 1968, {Atlas de galaxias australes} (Observatorio
Astronomico, Cordoba)

\bibitem[{{Spolaor} {et~al.}(2010){Spolaor}, {Kobayashi}, {Forbes}, {Couch}, \&
  {Hau}}]{spoletal10}
{Spolaor}, M., {Kobayashi}, C., {Forbes}, D.~A., {Couch}, W.~J., \& {Hau},
  G.~K.~T. 2010, \mnras, 408, 272

\bibitem[{{Thomas} \& {Davies}(2006)}]{thda06}
{Thomas}, D. \& {Davies}, R.~L. 2006, \mnras, 366, 510

\bibitem[{{Thomas} {et~al.}(2003){Thomas}, {Maraston}, \& {Bender}}]{thmabe03}
{Thomas}, D., {Maraston}, C., \& {Bender}, R. 2003, \mnras, 339, 897

\bibitem[{{Vazdekis} {et~al.}(2010){Vazdekis}, {S{\'a}nchez-Bl{\'a}zquez},
  {Falc{\'o}n-Barroso}, {Cenarro}, {Beasley}, {Cardiel}, {Gorgas}, \&
  {Peletier}}]{vazdekis2010}
{Vazdekis}, A., {S{\'a}nchez-Bl{\'a}zquez}, P., {Falc{\'o}n-Barroso}, J.,
  {et~al.} 2010, \mnras, 404, 1639

\bibitem[{{Worthey} {et~al.}(1994){Worthey}, {Faber}, {Gonzalez}, \&
  {Burstein}}]{wortetal94}
{Worthey}, G., {Faber}, S.~M., {Gonzalez}, J.~J., \& {Burstein}, D. 1994,
  \apjs, 94, 687

\bibitem[{{Worthey} \& {Ottaviani}(1997)}]{worott97}
{Worthey}, G. \& {Ottaviani}, D.~L. 1997, \apjs, 111, 377

\end{thebibliography}
\end{document}